\documentclass[12pt, draftclsnofoot, onecolumn]{IEEEtran}
\usepackage{geometry}
\ifCLASSINFOpdf

\else

\fi
\usepackage[cmex10]{amsmath}
\usepackage{amssymb}
\usepackage{cite}
\usepackage{graphicx}
\usepackage{array,color}
\usepackage{amsmath}
\usepackage{stfloats}
\usepackage{graphicx}
\usepackage{subfigure}
\usepackage{tabularx}
\usepackage{epsfig,epsf,color,balance,cite}
\usepackage{algorithmic}
\usepackage{algorithm}
\usepackage{bm}
\usepackage{textcomp}

\begin{document}

\title{Multicell MIMO Communications Relying on Intelligent Reflecting Surfaces}
\author{ Cunhua Pan, Hong Ren, Kezhi Wang,   Wei Xu, Maged Elkashlan,  Arumugam Nallanathan, \IEEEmembership{Fellow, IEEE},   and Lajos Hanzo, \IEEEmembership{Fellow, IEEE}
\thanks{C. Pan, H. Ren, M. Elkashlan and A. Nallanathan are with the School of Electronic Engineering and Computer Science at  Queen Mary University of London, London E1 4NS, U.K. (e-mail:\{c.pan, h.ren, maged.elkashlan, a.nallanathan\}@qmul.ac.uk). K. Wang is with Department of Computer and Information Sciences, Northumbria University, UK. (e-mail: kezhi.wang@northumbria.ac.uk).   W. Xu is with National Mobile Communications Research Laboratory, Southeast University, Nanjing 210096, China. (e-mail: wxu@seu.edu.cn).  L. Hanzo is with the School of Electronics and Computer Science, University of Southampton, Southampton, SO17 1BJ, U.K. (e-mail: lh@ecs.soton.ac.uk). }
}

\maketitle
\vspace{-1.9cm}
\begin{abstract}
Intelligent reflecting surfaces (IRSs) constitute a disruptive wireless communication technique capable of creating a controllable propagation  environment.  In this paper, we propose to invoke an IRS at the cell boundary of multiple cells to assist the downlink transmission to cell-edge users, whilst mitigating  the inter-cell interference, which is a crucial issue in multicell communication systems. We aim for maximizing the weighted sum rate (WSR) of all users through jointly optimizing the active precoding matrices at the base stations (BSs) and the phase shifts at the IRS subject to each BS's power constraint and unit modulus constraint. Both the BSs and the users are equipped with multiple antennas, which   enhances the spectral efficiency by exploiting the spatial multiplexing gain. Due to the non-convexity of the problem, we first reformulate it into an equivalent one, which is solved by using the  block coordinate descent (BCD) algorithm, where the precoding matrices and phase shifts are alternately optimized. The optimal precoding matrices can be obtained in closed form, when fixing the phase shifts. A pair of efficient algorithms are proposed for solving the phase shift optimization problem, namely the Majorization-Minimization (MM) Algorithm and the Complex Circle Manifold (CCM) Method. Both algorithms are guaranteed to converge to at least locally optimal solutions.  We also extend the proposed algorithms to the more general multiple-IRS and network MIMO scenarios.  Finally, our simulation results confirm the advantages of introducing IRSs in enhancing the cell-edge user performance.
\end{abstract}

\IEEEpeerreviewmaketitle
\section{Introduction}

Next-generation wireless communication systems are expected to provide a 1000-fold increase in the network capacity over the operational system for satisfying the ever-increasing demand for higher data rates   driven by  emerging applications such as augmented reality (AR) and virtual reality (VR). To achieve this goal,  promising techniques relying on massive multiple-input multiple-output (MIMO) solutions \cite{wence}, millimeter wave (mmWave) communications \cite{andrews2014will} and ultra-dense cloud radio access networks (UD-CRAN) have been advocated \cite{cunhuamaga,cpan2017,cpanTWC2019}. By deploying a massive number of antennas at the base station (BS) for transmission over the millimeter-wave (mm-wave) bands, significant spectral efficiency improvements can be achieved by exploiting the joint benefits of a high spatial multiplexing gain and high  bandwidth. However, escalating signal processing complexity, increased hardware costs as well as high power consumption are incurred by the associated high number of  radio frequency (RF) chains operating in a  high frequency band. These issues erode their practical benefits. Although the access points (AP) can be densely deployed in UD-CRAN systems for reducing the  distance between the users and the APs, the limited fronthaul capacity becomes their performance bottleneck. Furthermore, these techniques have to operate in the face of unfavourable electromagnetic wave propagation, improving  a high blockage probability.

As a remedy, intelligent reflecting surface (IRS) has been proposed as a revolutional technique of facilitating  both spectrum- and energy-efficient communications through reconfiguring the wireless propagation environment \cite{di2019smart,qingqing2019towards}. An IRS consists of a vast number of low-cost passive reflecting elements, each of which can independently adjust the phase shift of the signals incident upon it, and thus  collaboratively creating favourable wireless transmission channels  by innovatively harnessing the reflected signal. By properly tuning the phase shifts by using an IRS controller, the reflected signals can be added constructively at the desired receiver for enhancing the received signal power, whilst destructively superimposing them at the non-intended receivers for reducing the co-channel interference. Although passive reflecting surfaces have already been used in radar systems, the phase shifts of passive elements cannot be changed once they were fabricated, and they are unable to control the wireless propagation channels. Fortunately, due to the recent advance in micro-electromechanical systems (MEMS) and metamaterials \cite{cui2014coding}, the phase shifts can now be adjusted in real time, which results in near-instantaneously reconfigurable IRS possible. Although an IRS resembles the classic amplify-and-forward (AF) relay, the former has the advantage of lower power consumption, since it only reflects the signals passively without requiring active RF chains, while the latter necessitates active RF components for signal transmission. Hence, IRSs do  not impose additional thermal noise on the reflected signals.  Performance comparisons between AF relay and IRS were performed in \cite{bjornson2019intelligent,ntontin2019reconfigurable}.  Given the limited functionality of IRSs, their phase shifters can be fabricated in a compact form. Hence, each IRS accommodates a large number of phase shifters and provides  high beamforming gains. Furthermore, IRSs have the appealing advantages of light weight and small sizes, which can  be readily installed at buildings facades, on the room-ceilings, on lamp posts, on road signs, etc. IRSs can also be integrated into the existing communication systems at a modest modification.  However, to reap the aforementioned benefits promised of IRSs, the phase shifts  have to be appropriately   optimized along with the active beamforming weights at the BS. The main difficulty in optimizing the phase shifts is the non-convex unit modulus constraint imposed on the phase shifts. Although this kind of constraints have been studied both in hybrid digital/analog precoding \cite{el2014spatially,yu2016alternating}  and in constant-envelope precoding in massive MIMO systems \cite{Mohammed2013,mohammed2012single}, these studies were  only focused on the designs at the transmitter, which are not applicable for the joint active beamforming design of the BSs and of the passive beamforming design at the IRS.

\begin{figure}
\centering
\includegraphics[width=2.8in]{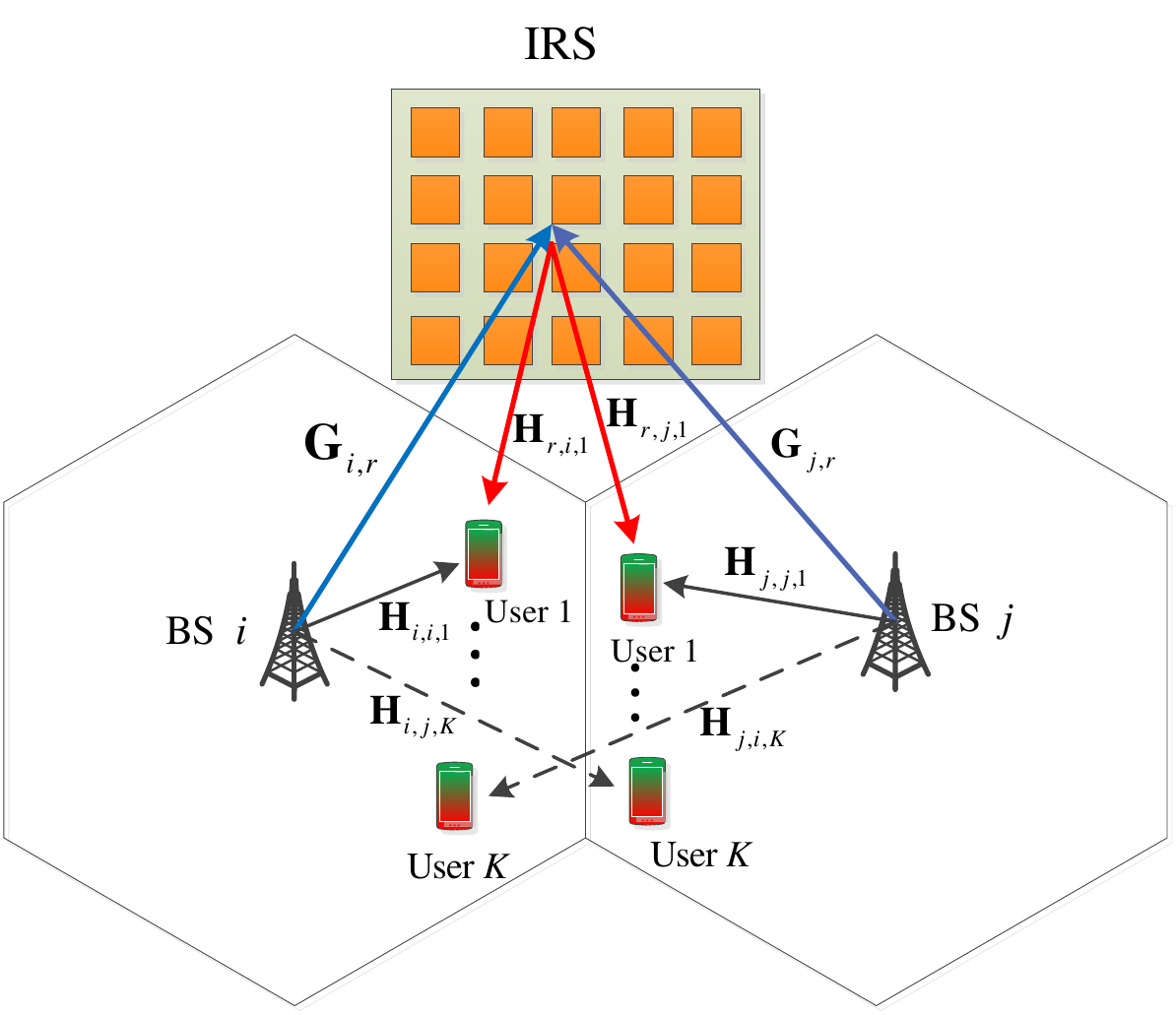}
\caption{An IRS-assisted multicell MIMO multiuser communication system.}\vspace{-0.8cm}
\label{fig1}
\end{figure}

Most recently,  some initial efforts have been devoted to the transmitter design of IRS-assisted wireless communication systems, including the single-user case of \cite{qingqingwuglobe,yu2019miso,yang2019intelligent,han2019large}, the downlink  multiuser case of \cite{wu2018intelligent,huang2019reconfigurable,guo2019weighted,nadeem2019large,taha2019enabling}, wireless power transfer design of \cite{cunhuasubjsac}, mobile edge computing of \cite{bai2019latency}, multicast scenario of \cite{zhou2019intelligent}  and the physical layer security design of \cite{yu2019enabling,cui2019secure,hongshen,chen2019intelligent,hong2020artificial}. However, the above-mentioned papers only studied the single-cell scenario, whilst there is a paucity of investigations on the multicell scenario  in the existing literature. To mitigate the spectrum scarcity, different cells will reuse the same frequency resources, which causes  severe inter-cell interference, especially for cell-edge users \cite{Gesbert}. Hence, in this paper, we propose to employ an IRS at the cell boundary for assisting the cell-edge users of multicell systems as shown in Fig.~\ref{fig1}, where the inter-cell interference can be alleviated with the aid of IRSs. Specifically, by carefully adjusting the phase shifts of the IRS's reflective elements, the inter-cell interference reflected by the IRS can be superimposed destructively on the direct interference impinging from the adjacent BS for minimizing the interference power at the receivers. This provides a higher degree of freedom for designing the beamforming/precoding at each BS for the users in its own cell. As a result, the active beamforming/precoding applied at each BS and the passive beamforming matrix of the IRS have to be jointly optimized. However, the resultant optimization problem is challenging to solve, since the optimization variables are highly coupled. Furthermore, all the existing contributions consider the single-antenna aided user scenario. However, owing to the rapid developments in antenna technology \cite{huang2014design}, the user equipment  is also capable of accommodating  multiple antennas for enhancing the received signal strength. Then, multiple data streams can be transmitted simultaneously, which boost the throughput. Therefore, in this paper, we consider the multiple-antenna aided user scenario. Given the complex mathematical data rate expression, the  techniques conceived in \cite{qingqingwuglobe,yu2019miso,yang2019intelligent,han2019large,wu2018intelligent,huang2019reconfigurable,guo2019weighted,nadeem2019large,taha2019enabling,yu2019enabling,cui2019secure,hongshen,chen2019intelligent}
cannot be directly applied. The multiple-antenna user case further complicates the optimization.

The main contributions of this paper can be summarized as follows:
\begin{enumerate}
  \item To the best of our knowledge, this is the first attempt to explore the assistance of IRSs in enhancing the cell-edge  performance in multicell MIMO communication systems. Specifically, we jointly optimize the active  transmit precoding (TPC) matrices of all BSs and the phase shifts at the IRS for maximizing the weighted sum rate (WSR) of all users subject to each BS's power constraint and to the unit modulus constraint of the phase shifters. However, the objective function (OF) is not jointly concave over both the TPC matrices and the phase shifts, which are highly coupled. To tackle this challenging problem, we first reformulate the original problem into an equivalent one by exploiting the  equivalence between the data rate and the weighted minimum  mean-square error (WMMSE). Then, the block coordinate descent (BCD) algorithm is proposed for alternately optimizing the TPC matrices at the BSs and the  passive beamforming at the IRS.
  \item Given the fixed phase shifts, we derive the optimal TPC matrices in closed form by applying the classic Lagrangian multiplier method. Since the phase shift optimization problem  is highly coupled with the various channel matrices and TPC matrices, this is quite a challenge. By using sophisticated matrix manipulations and transformations, we successfully transform the phase shift optimization problem into a non-convex quadratically constrained quadratic program (QCQP) subject to unit modulus constraint. A pair of efficient iterative algorithms are proposed for solving this problem. The first one is the Majorization-Minimization (MM) Algorithm \cite{yansun}, where a closed-form solution can be obtained in each iteration. The second is based on the Complex Circle Manifold (CCM) Method \cite{alhujaili2019transmit}, where  we show that the unit modulus constraints of all phase shifters constitute a complex circle manifold. Both the MM algorithm and the CCM algorithm are guaranteed to obtain at least a locally optimal solution.
  \item  The proposed algorithms are also extended to the more general multiple-IRS and network MIMO scenarios.
  \item Our simulation results show that the cell-edge performance can be significantly enhanced by employing   IRSs compared to a conventional multicell system operating without  IRSs. Moreover, it is also shown that the performance gain achieved by the IRS is indeed mainly due to the improving BS-IRS   and IRS-user links. Furthermore, the location of IRSs should be carefully chosen. It is shown that deploying IRSs at the cell boundary   achieves the highest gains for cell-edge users. Furthermore, simulation results also show that the IRSs should be deployed in the vicinity of the user clusters, and distributed IRS deployment has superior performance than the centralized deployment.
\end{enumerate}

The remainder of this paper is organized as follows. In Section \ref{system}, we present the system model of   IRS-assisted multicell MIMO communication and formulate the WSR maximization problem. In Section \ref{algo}, we reformulate the original problem into a more tractable problem and the TPC   matrices and passive beamforming phases are alternately optimized. In Section \ref{simlresult},  extensive simulation results are provided for quantifying  the performance advantages of introducing  IRSs into multicell systems.  Finally, our conclusions are offered in Section \ref{conclu}.

\emph{Notations}: For a complex value $a$, ${\rm{Re}}\{a\}$ represents the real part of $a$. Boldface lower case and
upper case letters denote vectors and matrices, respectively. ${{\mathbb{ C}}^{M }}$ denotes the set of $M \times 1$ complex vectors. ${{\mathbb{E}}}\{\cdot\} $ denotes the expectation operation. ${\left\| {\bf{x}} \right\|_2}$ denotes the 2-norm of vector ${\bf{x}}$. For  two matrices $\bf A$ and $\bf B$, ${\bf{A}} \odot {\bf{B}}$ represents the Hadamard product of $\bf A$ and $\bf B$. ${\left\| {\bf{A}} \right\|_F}$, ${\rm{Tr}}\left( {\bf{A}} \right)$ and $\left| {\bf{A}} \right|$  denote the  Frobenius norm, trace operation and determinant  of ${\bf{A}}$, respectively. $\nabla {f_{\bf{x}}}\left( {\bf{x}} \right)$ denotes the gradient of the function $f$ with respect to (w.r.t.) the vector ${\bf{x}} $. ${\cal C}{\cal N}({\bf{0}},{\bf{I}})$ represents a random vector following the distribution of zero mean and unit variance matrix. $\arg\{\cdot\}$ means the extraction of phase information. ${\rm{diag}}(\cdot)$ denotes the diagonalization operation. ${\left( \cdot \right)^{*}}$, ${\left( \cdot \right)^{\rm{T}}}$ and ${\left(  \cdot \right)^{\rm{H}}}$ denote the conjugate, transpose and Hermitian operators, respectively.

\section{System Model and Problem Formulation}\label{system}
\subsection{System  Model}

We consider an IRS-aided multicell downlink  MIMO   model constituted by $L$ macro cells, each of which has a single base station (BS) that serves $K$ cell-edge users.  Each BS and each user is equipped with $N_{t}\ge 1$ and $N_{r}\ge 1$ transmit antennas (TAs) and receive antennas (RAs), respectively. Each cell-edge user  suffers both from high   attenuation from its serving BS and severe cochannel interference from its neighbouring BSs. To mitigate these, we propose to employ an IRS which has $M$ reflection elements at the cell edge as shown in Fig. \ref{fig1}, which boost the useful signal power and mitigate the cochannel interference by carefully designing the  phase shifts of the reflective elements.

The signal  transmitted by the $l$th BS is given by
\begin{equation}\label{reghtjui}
  {{\bf{x}}_l} = \sum\limits_{k = 1}^K {{\bf{F}}_{l,k}{\bf{s}}_{l,k}},
\end{equation}
where ${\bf{s}}_{l,k}$ is the $(d \times 1)$-element  symbol vector transmitted to the $k$th user in its cell, satisfying $\mathbb{E}\left[ {{{\bf{s}}_{l,k}}{{{\bf{s}}_{l,k}^H} }} \right] = {\bf{I}}_d$ and $\mathbb{E}\left[ {{{\bf{s}}_{l,k}}{\left( {{\bf{s}}_{i,j}} \right)^H}} \right] = {\bf{0}}, {\rm{for}}\  \{l,k\} \ne \{i,j\}$, and ${\bf{F}}_{l,k}\in \mathbb{C}^{N_{t}\times d}$ is the linear TPC matrix used by the $l$th BS for transmitting its data vector ${\bf{s}}_{l,k}$ to the $k$th user. The baseband channels spanning from the $n$th BS to the $k$th user in the $l$th cell, as well as those from the IRS to the $k$th user in the $l$th cell, and the ones from the $n$th BS to the IRS are denoted by ${\bf{H}}_{n,l,k}$, ${\bf{H}}_{l,k}^{r}$ and ${\bf{G}}_{n}^{r}$, respectively. Let us denote the phase shift of the $m$-th reflection element of the IRS by ${\theta _m} \in [0,2\pi ]$. Thus the reflection operator simply multiplies the incident multi-path signals  by
${e^{j{\theta _m}}}$ \footnote{$j$ is the imaginary unit.} at a single physical point and then forwards the combined signal to the users. Hence, the users will directly receive the desired signals from the BSs, plus the signals reflected by the IRS. However, we ignore the signal  reflected more than once due to the severe path loss. Let us  denote the diagonal phase-shifting matrix of the IRS as ${\bm{\Phi}}  = {\rm{diag}}\left\{ {{e^{j{\theta _1}}}, \cdots ,{e^{j{\theta _m}}}, \cdots ,{e^{j{\theta _M}}}} \right\}$. Then, the received signal vector at the $k$th user in the $l$th cell is given by
\begin{equation}\label{uhuhhu}
{{\bf{y}}_{l,k}} = \underbrace {\sum\limits_{n = 1}^L {{{\bf{H}}_{n,l,k}}{{\bf{x}}_n}} }_{{\rm{Siganls\  from\  BSs}}} + \underbrace {\sum\limits_{n = 1}^L {{{\bf{H}}_{l,k}^{r}}{\bm{\Phi}} {{\bf{G}}_{n}^{r}}} {{\bf{x}}_n}}_{{\rm{Signals\  from\  the\  IRS}}} + {{\bf{n}}_{l,k}},
\end{equation}
where  ${\bf{n}}_{l,k}$ is the noise vector that satisfies ${\cal C}{\cal N}\left( {{\bf{0}},\sigma^2{{\bf{I}}_{N_r}}} \right)$.

We assume that the channel state information (CSI) of all channels is perfectly known at the BS, and the BS calculates the optimal phase shifts and sends them back to the IRS controller. Indeed,  the assumption of having perfect CSI knowledge at the BS is idealistic because it is challenging to obtain the CSI in IRS-assisted communication systems. However, the algorithms developed allow  us to derive   the relevant performance upper bounds for realistic scenarios in the presence of realistic CSI errors. In addition, the proposed algorithms can provide  insights into the performance gain provided by  IRSs, which can inspire  further research  in this area. Recently, we have conceived a framework for the robust transmission design of an IRS-aided single-cell scenario \cite{zhou2020framework} by considering both the bounded CSI error model and the statistical CSI error model associated with the cascaded channels. Its extension to the multicell scenario will be studied in our future research.

Let us define  ${\bf{\bar H}}_{n,l,k} \buildrel \Delta \over =  {\bf{H}}_{l,k}^{r}{\bm{\Phi}} {{\bf{G}}_{n}^{r}} + {\bf{H}}_{n,l,k}$, which can be regarded as the equivalent channel spanning from the $n$th BS to the $k$th user in the $l$th cell. By substituting (\ref{reghtjui}) into (\ref{uhuhhu}), ${\bf{y}}_{l,k} $ can be written as
\begin{equation}\label{ijoiotr}
 {\bf{y}}_{l,k} = {\bf{\bar H}}_{l,l,k}{\bf{F}}_{l,k}{\bf{s}}_{l,k} + \underbrace {\sum\limits_{m = 1,m \ne k}^K {{\bf{\bar H}}_{l,l,k}{\bf{F}}_{l,m}{\bf{s}}_{l,m}} }_{{\rm{Intra - cell interference}}} + \underbrace {\sum\limits_{n = 1,n \ne l}^L {\sum\limits_{m = 1}^K {{\bf{\bar H}}_{n,l,k}{\bf{F}}_{n,m}{\bf{s}}_{n,m}} } }_{{\rm{Inter - cell interference}}} + {\bf{n}}_{l,k}.
\end{equation}
Then, the achievable data rate (nat/s/Hz) of the $k$th user in the $l$th cell is given by \cite{cpan2017}
\begin{equation}\label{hufureu}
{R_{l,k}}\left( {{\bf{F}},\bm{\theta} } \right) = {\log}\left| {{\bf{I}} + {{{\bf{\bar H}}}_{l,l,k}}{{\bf{F}}_{l,k}}{\bf{F}}_{l,k}^{\rm{H}}{\bf{\bar H}}_{l,l,k}^{\rm{H}}{\bf{J}}_{l,k}^{ - 1}} \right|,
\end{equation}
where we have ${\bf{F}} = \left[ {{{\bf{F}}_{l,k}},\forall l,k} \right],{\bm{\theta}}  = \left[ {{\theta _1}, \cdots ,{\theta _M}} \right]$, and ${{\bf{J}}_{l,k}}$ is the interference-plus-noise covariance matrix given by
\begin{equation}\label{ohhugth}
 {{\bf{J}}_{l,k}} = \sum\limits_{m = 1,m \ne k}^K {{{{\bf{\bar H}}}_{l,l,k}}{{\bf{F}}_{l,m}}{\bf{F}}_{l,m}^{\rm{H}}{\bf{\bar H}}_{l,l,k}^{\rm{H}}}  + \sum\limits_{n = 1,n \ne l}^L {\sum\limits_{m = 1}^K {{{{\bf{\bar H}}}_{n,l,k}}{{\bf{F}}_{n,m}}{\bf{F}}_{n,m}^{\rm{H}}{\bf{\bar H}}_{n,l,k}^{\rm{H}}} }  + {\sigma ^2}{\bf{I}}.
\end{equation}

\subsection{Problem Formulation}

In this paper, we aim for maximizing the WSR of all the users by jointly optimizing the TPC matrices ${\bf{F}}$ at the BSs and the phase shifts ${\bm{\theta}}$ at the IRS, while guaranteeing the total power constraint at each BS. Specifically, the WSR maximization problem is formulated as:
\begin{subequations}\label{appstaoneorig}
\begin{align}
\mathop {\max }\limits_{{{\bf{F}},{\bm{\theta}} } } \quad
& \sum\limits_{l = 1}^L {\sum\limits_{k = 1}^K {{\omega_{l,k}}{R_{l,k}}\left( {{\bf{F}},{\bm{\theta}} } \right)} }
\\
\qquad\ \textrm{s.t.}\qquad
&\sum\limits_{k = 1}^K {\left\| {{{\bf{F}}_{l,k}}} \right\|_F^2}  \le {P_{l,\max }},l = 1, \cdots ,L,\label{dehwifr}\\
& 0 \le {\theta _m} \le 2\pi, m = 1, \cdots ,M, \label{jiofjj}
\end{align}
\end{subequations}
where $\omega_{l,k}$ denotes the weighting factor   representing the priority of the corresponding user. Due to the coupling effect between the TPC matrices ${\bf{F}}$ and the phase shifts ${\bm{\theta}}$, this optimization problem is   difficult to solve.   Additionally, the phase shift constraints in (\ref{jiofjj}) further aggravate the challenge.
In the following, we provide a low-complexity algorithm for solving Problem (\ref{appstaoneorig}).

 \section{Low-Complexity Algorithm Development }\label{algo}
In this section, we first reformulate the original problem into a more tractable form. Then, the block coordinate
descent (BCD) method is proposed for solving the formulated problem.
\subsection{Reformulation of the Original Problem}
In the following, we exploit the relationship between the data rate and the mean-square error (MSE) for the optimal decoding matrix.  To reduce the decoding complexity, we consider a linear decoding matrix so that the estimated signal vector of each user is given by
\begin{equation}\label{edwfre}
 {{{\bf{\hat s}}}_{l,k}} = {\bf{U}}_{l,k}^{\rm{H}}{{\bf{y}}_{l,k}},\forall l,k,
\end{equation}
where ${\bf{U}}_{l,k}\in \mathbb{C}^{N_r \times d}$ is the decoding matrix for the $k$th user in the $l$th cell. Then, the MSE matrix of each user is given by
\begin{eqnarray}
  {{\bf{E}}_{l,k}} &=&{\mathbb{E}_{{\bf{s,n}}}}\left[ {\left( {{{{\bf{\hat s}}}_{l,k}} - {{\bf{s}}_{l,k}}} \right){{\left( {{{{\bf{\hat s}}}_{l,k}} - {{\bf{s}}_{l,k}}} \right)}^H}} \right]\\
   &=& \left( {{\bf{U}}_{l,k}^{\rm{H}}{{{\bf{\bar H}}}_{l,l,k}}{{\bf{F}}_{l,k}} - {\bf{I}}} \right){\left( {{\bf{U}}_{l,k}^{\rm{H}}{{{\bf{\bar H}}}_{l,l,k}}{{\bf{F}}_{l,k}} - {\bf{I}}} \right)^{\rm{H}}} + \sum\limits_{m = 1,m \ne k}^K {{\bf{U}}_{l,k}^{\rm{H}}{{{\bf{\bar H}}}_{l,l,k}}{{\bf{F}}_{l,m}}{\bf{F}}_{l,m}^{\rm{H}}{\bf{\bar H}}_{l,l,k}^{\rm{H}}{{\bf{U}}_{l,k}}} \nonumber\\
  && + \sum\limits_{n = 1,n \ne l}^L {\sum\limits_{m = 1}^K {{\bf{U}}_{l,k}^{\rm{H}}{{{\bf{\bar H}}}_{n,l,k}}{{\bf{F}}_{n,m}}{\bf{F}}_{n,m}^{\rm{H}}{\bf{\bar H}}_{n,l,k}^{\rm{H}}{{\bf{U}}_{l,k}}} }  + {\sigma ^2}{\bf{U}}_{l,k}^{\rm{H}}{{\bf{U}}_{l,k}}, \forall l,k.\label{jvjgjgreji}
\end{eqnarray}

Upon introducing a set of auxiliary matrices ${\bf{W}}=\{{{\bf{W}}_{l,k}}\succeq {\bf{0}}, \forall l,k\}$ and defining ${\bf{U}}=\{{{\bf{U}}_{l,k}}, \forall l,k\}$, Problem (\ref{appstaoneorig}) can be reformulated as follows \cite{shi2011iteratively,cpan2017}:
 \begin{subequations}\label{appstneorig}
\begin{align}
\mathop {\max }\limits_{{{\bf{W}}, {\bf{U}}, {\bf{F}},{\bm{\theta}} } } \quad
& \sum\limits_{l = 1}^L {\sum\limits_{k = 1}^K {{\omega_{l,k}}{h_{l,k}}\left( {{\bf{W}}, {\bf{U}},{\bf{F}},{\bm{\theta}} } \right)} }
\\
\qquad\ \textrm{s.t.}\qquad
&\sum\limits_{k = 1}^K {\left\| {{{\bf{F}}_{l,k}}} \right\|_F^2}  \le {P_{l,\max }},l = 1, \cdots ,L,\label{defr}\\
& 0 \le {\theta _m} \le 2\pi, m = 1, \cdots ,M, \label{ofjj}
\end{align}
\end{subequations}
where  ${h_{l,k}}\left( {{\bf{W}}, {\bf{U}},{\bf{F}},{\bm{\theta}} } \right)$ is given by
\begin{equation}\label{fdgtshdy}
  {h_{l,k}}\left( {{\bf{W}}, {\bf{U}},{\bf{F}},{\bm{\theta}} } \right)={\log \left| {{{\bf{W}}_{l,k}}} \right| - {\rm{Tr}}\left( {{{\bf{W}}_{l,k}}{{\bf{E}}_{l,k}}} \right) + d}.
\end{equation}
Note that compared to the original OF of Problem (\ref{appstaoneorig}), the new OF in Problem (\ref{appstneorig}) is in a more tractable form, although we have introduced more optimization variables. For a given phase shift ${\bm{\theta}}$, ${h_{l,k}}\left( {{\bf{W}}, {\bf{U}},{\bf{F}},\bm{\theta} } \right)$ is a concave function for each set of the optimization matrices, when the other two are fixed. In the following, we propose the BCD algorithm for solving Problem (\ref{appstneorig}). Specifically, we maximize the OF in (\ref{appstneorig}) by alternately optimizing one set of optimization variables, while keeping the other variables fixed. Note that the decoding matrix ${\bf{U}}_{l,k}$ and the auxiliary matrix ${\bf{W}}_{l,k}$ are only related to ${h_{l,k}}\left( {{\bf{W}}, {\bf{U}},{\bf{F}}, \bm{\theta} } \right)$. In the following, we can derive the optimal solution for ${\bf{U}}_{l,k}$ and ${\bf{W}}_{l,k}$, when the other matrices are fixed.
For given values of ${\bm{\theta}}$, ${\bf{W}}$, and ${\bf{F}}$, we can set the first-order derivative of ${h_{l,k}}\left( {{\bf{W}}, {\bf{U}},{\bf{F}}, \bm{\theta} } \right)$ with respect to ${\bf{U}}_{l,k}$   to zero, which gives the optimal ${\bf{U}}_{l,k}$:
\begin{equation}\label{fgrtgtyh}
 {\bf{U}}_{l,k} = {\left( {{{\bf{J}}_{l,k}} + {{{\bf{\bar H}}}_{l,l,k}}{{\bf{F}}_{l,k}}{\bf{F}}_{l,k}^{\rm{H}}{\bf{\bar H}}_{l,l,k}^{\rm{H}}} \right)^{ - 1}}{{{\bf{\bar H}}}_{l,l,k}}{{\bf{F}}_{l,k}}.
\end{equation}
Similarly, for given ${\bm{\theta}}$, ${\bf{U}}$, and ${\bf{F}}$, the optimal auxiliary matrix ${\bf{W}}_{l,k}$ can be obtained as follows:
\begin{equation}\label{bhuj}
 {\bf{W}}_{l,k} = {\bf{E}}_{l,k}^{ - 1},
\end{equation}
where ${\bf{E}}_{l,k}$ is given in (\ref{jvjgjgreji}).

Let us now focus our attention on optimizing the TPC matrices ${\bf{F}}$ and phase shifts ${\bm{\theta}}$.

\subsection{Optimizing the Precoding Matrices ${\bf{F}}$}\label{kodsijcosakpdc}

In this subsection, we focus our attention on optimizing the TPC matrices ${\bf{F}}$, while fixing ${\bf{W}}, {\bf{U}}$ and $\bm{\theta}$. By substituting ${\bf{E}}_{l,k}$ into (\ref{fdgtshdy}), the optimization over ${\bf{F}}$ can be decoupled among the different BSs. Specifically, by removing the constant terms, the TPC matrix optimization problem of the $l$th BS is given by
 \begin{subequations}\label{appssxsorig}
\begin{align}
&
\begin{array}{l}
\mathop {\min }\limits_{{ {\bf{F}}_{l,k},\forall k } }\quad   \sum\limits_{n = 1}^L {\sum\limits_{m = 1}^K {{\omega _{n,m}}{\rm{Tr}}} } \left( {{{\bf{W}}_{n,m}}\sum\limits_{k = 1}^K {{\bf{U}}_{n,m}^{\rm{H}}{{{\bf{\bar H}}}_{l,n,m}}{{\bf{F}}_{l,k}}{\bf{F}}_{l,k}^{\rm{H}}{\bf{\bar H}}_{l,n,m}^{\rm{H}}{{\bf{U}}_{n,m}}} } \right)\\
 \qquad\quad - \sum\limits_{k = 1}^K {{\omega _{l,k}}{\rm{Tr}}\left( {{{\bf{W}}_{l,k}}{\bf{U}}_{l,k}^{\rm{H}}{{{\bf{\bar H}}}_{l,l,k}}{{\bf{F}}_{l,k}}} \right)}  - \sum\limits_{k = 1}^K {{\omega _{l,k}}{\rm{Tr}}\left( {{{\bf{W}}_{l,k}}{\bf{F}}_{l,k}^{\rm{H}}{\bf{\bar H}}_{l,l,k}^{\rm{H}}{{\bf{U}}_{l,k}}} \right)}
\end{array}
\\
&\textrm{s.t.}\quad \sum\limits_{k = 1}^K {\left\| {{{\bf{F}}_{l,k}}} \right\|_F^2}  \le {P_{l,\max }}.\label{djjkfr}
\end{align}
\end{subequations}
It can be readily verified that the above  problem is a convex optimization problem, which can be transformed into a second order cone programming (SOCP) problem  that can be efficiently solved by using  standard optimization packages, such as CVX \cite{grant2014cvx}. However, the computational complexity of solving an SOCP problem is   high. To reduce the complexity, in the following we provide a near-optimal closed-form expression of the TPC matrices by using the Lagrangian multiplier method.

Following some further manipulations, the Lagrangian function of Problem (\ref{appssxsorig})  is written as
\begin{equation}\label{rfgt}
  \begin{array}{l}
{\cal L}\left( {{{\bf{F}}_{l,k}},\forall k,{\lambda _l}} \right) = \sum\limits_{k = 1}^K {{\rm{Tr}}\left( {{\bf{F}}_{l,k}^{\rm{H}}\left( {{{\bf{A}}_{l}} + {\lambda _l}{\bf{I}}} \right){{\bf{F}}_{l,k}}} \right)}  - \sum\limits_{k = 1}^K {{\omega _{l,k}}{\rm{Tr}}\left( {{{\bf{W}}_{l,k}}{\bf{U}}_{l,k}^{\rm{H}}{{{\bf{\bar H}}}_{l,l,k}}{{\bf{F}}_{l,k}}} \right)} \\
 \qquad \qquad\qquad\quad - \sum\limits_{k = 1}^K {{\omega _{l,k}}{\rm{Tr}}\left( {{{\bf{W}}_{l,k}}{\bf{F}}_{l,k}^{\rm{H}}{\bf{\bar H}}_{l,l,k}^{\rm{H}}{{\bf{U}}_{l,k}}} \right)}  - {\lambda _l}{P_{l,\max }},
\end{array}
\end{equation}
where ${\lambda _l}\ge 0$ is the Lagrangian multiplier associated with the power constraint of the $l$th BS, and ${{{\bf{A}}_{l,k}}}$ is given by
\begin{equation}\label{wsdwe}
  {{\bf{A}}_l} = \sum\limits_{n = 1}^L {\sum\limits_{m = 1}^K {{\omega _{n,m}}{\bf{\bar H}}_{l,n,m}^{\rm{H}}{{\bf{U}}_{n,m}}{{\bf{W}}_{n,m}}{\bf{U}}_{n,m}^{\rm{H}}{{{\bf{\bar H}}}_{l,n,m}}} }.
\end{equation}
By setting the first-order derivative of  ${\cal L}\left( {{{\bf{F}}_{l,k}},\forall k,{\lambda _l}} \right)$ w.r.t. ${{{\bf{F}}_{l,k}}}$ to zero, we can obtain the optimal solution of ${\bf{F}}_{l,k}$ as follows:
\begin{equation}\label{dwfohu}
  {{\bf{F}}_{l,k}}({\lambda _l}) = {\omega _{l,k}}{\left( {{{\bf{A}}_l} + {\lambda _l}{\bf{I}}} \right)^{ \dag }}{\bf{\bar H}}_{l,l,k}^{\rm{H}}{{\bf{U}}_{l,k}}{{\bf{W}}_{l,k}},
\end{equation}
where ${(\cdot)^\dag }$ denotes the matrix pseudoinverse. The value of ${\lambda _l}$ should be chosen for ensuring that the following complementary slackness condition for the power constraint is satisfied:
\begin{equation}\label{aswdewf}
{\lambda _l}\left( {\sum\limits_{k = 1}^K {\left\| {{{\bf{F}}_{l,k}}} ({\lambda _l})\right\|_F^2}  - {P_{l,\max }}} \right) = 0.
\end{equation}
In the following, we elaborate on how to obtain the optimal ${\lambda _l}$, which is divided into two cases: 1) ${{\bf{A}}_l}$ is full rank; 2) ${{\bf{A}}_l}$ is low rank.

\subsubsection{Case I: ${{\bf{A}}_l}$ is full rank}

In this case, ${{\bf{A}}_l}$ is a positive definite matrix, which can be decomposed as ${{\bf{A}}_l} = {\bf{Q}}_l{\bm{\Lambda}}_l {{\bf{Q}}_l^{\rm{H}}}$ by using the singular value decomposition (SVD), where ${\bf{Q}}_l{{\bf{Q}}_l^{\rm{H}}} = {{\bf{Q}}_l^{\rm{H}}}{\bf{Q}}_l = {{\bf{I}}_{{N_t}}}$ and ${\bm{\Lambda}}_l$ is a diagonal matrix with positive diagonal elements. Then, we have
\begin{eqnarray}
\!\!\!  f_l({\lambda _l})&\buildrel \Delta \over =&\sum\limits_{k = 1}^K {{\rm{Tr}}\left( {{\bf{F}}_{l,k}({\lambda _l})^{\rm{H}}{{\bf{F}}_{l,k}({\lambda _l})}} \right)} \nonumber\\
   &=& \sum\limits_{k = 1}^L {\omega _{l,k}^2{\rm{Tr}}\left( {{\bf{W}}_{l,k}^{\rm{H}}{\bf{U}}_{l,k}^{\rm{H}}{{{\bf{\bar H}}}_{l,l,k}}{{\left( {{{\bf{A}}_l} + {\lambda _l}{\bf{I}}} \right)}^{ - 1}}{{\left( {{{\bf{A}}_l} + {\lambda _l}{\bf{I}}} \right)}^{ - 1}}{\bf{\bar H}}_{l,l,k}^{\rm{H}}{{\bf{U}}_{l,k}}{{\bf{W}}_{l,k}}} \right)} \label{ifuhio} \\
   &=& \sum\limits_{k = 1}^L {\omega _{l,k}^2{\rm{Tr}}\left( {{\bf{W}}_{l,k}^{\rm{H}}{\bf{U}}_{l,k}^{\rm{H}}{{{\bf{\bar H}}}_{l,l,k}}{{\left( {{\bf{Q}}_l{\bm{\Lambda}}_l {{\bf{Q}}_l^{\rm{H}}} + {\lambda _l}{\bf{I}}} \right)}^{ - 1}}{{\left( {{\bf{Q}}_l{\bm{\Lambda}}_l {{\bf{Q}}_l^{\rm{H}}} + {\lambda _l}{\bf{I}}} \right)}^{ - 1}}{\bf{\bar H}}_{l,l,k}^{\rm{H}}{{\bf{U}}_{l,k}}{{\bf{W}}_{l,k}}} \right)}  \\
   &=& \sum\limits_{k = 1}^L {\omega _{l,k}^2{\rm{Tr}}\left( {{\bf{W}}_{l,k}^{\rm{H}}{\bf{U}}_{l,k}^{\rm{H}}{{{\bf{\bar H}}}_{l,l,k}}{\bf{Q}}_l{{\left( {{\bm{\Lambda}}_l  + {\lambda _l}{\bf{I}}} \right)}^{ - 2}}{{\bf{Q}}_l^{\rm{H}}}{\bf{\bar H}}_{l,l,k}^{\rm{H}}{{\bf{U}}_{l,k}}{{\bf{W}}_{l,k}}} \right)} \label{jufe} \\
   &=&  {{\rm{Tr}}\left( {{{\left( {{\bm{\Lambda}}_l + {\lambda _l}{\bf{I}}} \right)}^{ - 2}}{{\bf{Z}}_{l}}} \right)}\label{jiehae}\\
   &=&\sum\limits_{i = 1}^{{N_t}} {\frac{{{{\left[ {{{\bf{Z}}_l}} \right]}_{i,i}}}}{{{{\left( {{{\left[ {{\bm{\Lambda}}_l} \right]}_{i,i}} + {\lambda _l}} \right)}^2}}}}, \label{kojmg}
\end{eqnarray}
where ${{\bf{Z}}_{l}} = \sum\limits_{k = 1}^L \omega _{l,k}^2{\bf{Q}}_l^{\rm{H}}{\bf{\bar H}}_{l,l,k}^{\rm{H}}{{\bf{U}}_{l,k}}{{\bf{W}}_{l,k}}{\bf{W}}_{l,k}^{\rm{H}}{\bf{U}}_{l,k}^{\rm{H}}{{{\bf{\bar H}}}_{l,l,k}}{{\bf{Q}}_l}$, ${\left[ {{{\bf{Z}}_l}} \right]}_{i,i}$ and ${\left[ {{\bm{\Lambda}}_l} \right]}_{i,i}$ denote the $i$th diagonal element of matrix ${{{\bf{Z}}_l}}$ and matrix ${\bm{\Lambda}}_l$, respectively. It can be readily verified that $f_l({\lambda _l})$ is a monotonically decreasing function. Hence, if $f_l(0)\le {P_{l,\max }}$, then the optimal TPC matrix is given by ${{\bf{F}}_{l,k}^{\rm{opt}}}={{\bf{F}}_{l,k}}(0)$. Otherwise, the optimal $\lambda_l$ can be obtained by using the bisection based search method  to find the solution of the following equation:
\begin{equation}\label{dergt}
f_l({\lambda _l})=\sum\limits_{i = 1}^{{N_t}} {\frac{{{{\left[ {{{\bf{Z}}_l}} \right]}_{i,i}}}}{{{{\left( {{{\left[ {{\bm{\Lambda}}_l} \right]}_{i,i}} + {\lambda _l}} \right)}^2}}}}={P_{l,\max }}.
\end{equation}
Since $f_l(\infty )=0$, the solution of Equation (\ref{dergt}) must exist, which is denoted as $\lambda _l^{\rm{opt}}$. Then, the optimal TPC matrix can be obtained as ${{\bf{F}}_{l,k}^{\rm{opt}}}={{\bf{F}}_{l,k}}(\lambda _l^{\rm{opt}})$. To apply the bisection based search method, we have to find the upper bound of $\lambda _l$, which is given by
\begin{equation}\label{xddfcerf}
  {\lambda _l} < \sqrt {\frac{{\sum\limits_{i = 1}^{{N_t}} {{{\left[ {{{\bf{Z}}_l}} \right]}_{i,i}}} }}{{{P_{l,\max }}}}}  \buildrel \Delta \over = \lambda _l^{{\rm{ub}}}.
\end{equation}
This can be proved as follows:
\begin{equation}\label{asdftg}
 {f_l}(\lambda _l^{{\rm{ub}}})=\sum\limits_{i = 1}^{{N_t}} {\frac{{{{\left[ {{{\bf{Z}}_l}} \right]}_{i,i}}}}{{{{\left( {{{\left[ {{\bm{\Lambda}}_l} \right]}_{i,i}} + {\lambda _l^{{\rm{ub}}}}} \right)}^2}}}} < \sum\limits_{i = 1}^{{N_t}} {\frac{{{{\left[ {{{\bf{Z}}_l}} \right]}_{i,i}}}}{{{{\left( {\lambda _l^{{\rm{ub}}}} \right)}^2}}}}  = {P_{l,\max }}.
\end{equation}

\subsubsection{Case II: ${{\bf{A}}_l}$ is low rank}

In this case, the above method cannot be directly applied since the ${\bf{Q}}_l$ obtained by SVD is not a unitary matrix, hence the step in (\ref{jufe}) cannot be applied. To resolve this issue, we first check whether $\lambda _l=0$ is the optimal solution or not.  If
\begin{equation}\label{ewfrg}
  f_l({0})=\sum\limits_{k = 1}^K {{\rm{Tr}}\left( {{\bf{F}}_{l,k}(0)^{\rm{H}}{{\bf{F}}_{l,k}}}(0) \right)}\le {P_{l,\max }},
\end{equation}
then the optimal TPC matrix is given by ${{\bf{F}}_{l,k}^{\rm{opt}}}={{\bf{F}}_{l,k}}(0)$, otherwise, the optimal $\lambda _l$ is a positive value, which will be obtained as follows. Upon defining the rank of ${{\bf{A}}_l}$ as $r_l={\rm{rank}}({\bf{A}}_l)<N_t$ and using the SVD, we have
\begin{equation}\label{xafrrf}
  {{\bf{A}}_l} = \left[ {{{\bf{Q}}_{l,1}},{{\bf{Q}}_{l,2}}} \right]{{\bm\Lambda} _l}{\left[ {{{\bf{Q}}_{l,1}},{{\bf{Q}}_{l,2}}} \right]^{\rm{H}}},
\end{equation}
where ${\bf{Q}}_{l,1}$ contains the first $r_l$ singular vectors corresponding to
the $r_l$ positive eigenvalues, and ${\bf{Q}}_{l,2}$ holds the last $N_t-r_l$ singular vectors corresponding to the $N_t-r_l$ zero-valued eigenvalues, ${{\bm{\Lambda}} _l} = {\rm{diag}}\left\{ {{{\bm{\Lambda}} _{l,1}},{{\bf{0}}_{\left( {{N_t} - {r_l}} \right) \times \left( {{N_t} - {r_l}} \right)}}} \right\}$ with ${\bm{\Lambda}} _{l,1}$ denoting the diagonal matrix containing the first $r_l$ positive eigenvalues. Upon defining ${{\bf{Q}}_l} \buildrel \Delta \over =  \left[ {{{\bf{Q}}_{l,1}},{{\bf{Q}}_{l,2}}} \right]$ and applying   similar steps to those in  (\ref{ifuhio}) to (\ref{kojmg}), we have
\begin{eqnarray}
  f_l({\lambda _l})&  =&\sum\limits_{k = 1}^K {{\rm{Tr}}\left( {{\bf{F}}_{l,k}({\lambda _l})^{\rm{H}}{{\bf{F}}_{l,k}({\lambda _l})}} \right)} \\
    &=& \sum\limits_{i = 1}^{{r_l}} {\frac{{{{\left[ {{{\bf{Z}}_l}} \right]}_{i,i}}}}{{{{\left( {{{\left[ {{{\bm{\Lambda}} _l}} \right]}_{i,i}} + {\lambda _l}} \right)}^2}}}}  + \sum\limits_{i = {r_l} + 1}^{{N_t}} {\frac{{{{\left[ {{{\bf{Z}}_l}} \right]}_{i,i}}}}{{\lambda _l^2}}},
\end{eqnarray}
where ${\bf{Z}}_l$ is the same as that in Case I. It is plausible that $f_l({\lambda _l})$ is a monotonically decreasing function for $\lambda _l>0$ and the optimal $\lambda _l$ can be obtained by using the bisection based search method, where the lower bound of $\lambda _l$ is set to a  small positive value.

 The overall algorithm to solve  Problem (\ref{appssxsorig}) is summarized in Algorithm \ref{iteda}.
\begin{algorithm}
\caption{ Bisection Search Method to Solve Problem (\ref{appssxsorig}) }\label{iteda}
\begin{algorithmic}[1]
\STATE  Initialize  the accuracy $\varepsilon$, the bounds $\lambda_l^{\rm{lb}}$ and $\lambda_l^{\rm{ub}}$;
\STATE  If $f_l({0})\le {P_{l,\max }}$ holds, the optimal TPC matrix is given by ${{\bf{F}}_{l,k}^{\rm{opt}}}={{\bf{F}}_{l,k}}(0), \forall k$ and terminate; Otherwise, go to step 3;
\STATE  Calculate  $\lambda_l  = {{\left( {\lambda_l^{\rm{lb}} + \lambda_l^{\rm{ub}}} \right)} \mathord{\left/
 {\vphantom {{\left( {{\lambda _l} + {\lambda _u}} \right)} 2}} \right.
 \kern-\nulldelimiterspace} 2}$;
 \STATE   If $f_l({\lambda_l})\le {P_{l,\max }}$, set ${\lambda _l^{\rm{ub}}}={\lambda_l}$. Otherwise, set ${\lambda _l^{\rm{lb}}}={\lambda_l}$;
 \STATE   If  $\left| {\lambda_l^{\rm{lb}} - \lambda_l^{\rm{ub}}} \right| \le \varepsilon $, terminate. Otherwise, go to step 2.
\end{algorithmic}
\end{algorithm}

\subsection{Optimizing the Phase Shifts ${\bm{\theta}}$}\label{hwdi}

In this subsection, we focus our attention on optimizing the phase shifts ${\bm{\theta}}$, while fixing ${\bf{W}}, {\bf{U}}$ and ${\bf{F}}$. By substituting ${\bf{E}}_{l,k}$ into (\ref{fdgtshdy}) and ignoring the terms that are not related to the channels, the phase shift optimization problem is formulated as:
 \begin{subequations}\label{appsdsworig}
\begin{align}
&\begin{array}{*{20}{l}}
{\mathop {\min }\limits_{\bm{\theta}}  \quad \sum\limits_{l = 1}^L {\sum\limits_{n = 1}^L \sum\limits_{m = 1}^K {{\rm{Tr}}}  \left( {\omega _{n,m}}{{{\bf{W}}_{n,m}}{\bf{U}}_{n,m}^{\rm{H}}{{{\bf{\bar H}}}_{l,n,m}}{\bf{F}}_l {{\bf{\bar H}}_{l,n,m}^{\rm{H}}{{\bf{U}}_{n,m}}} } \right)} }\\
{\qquad \quad  - \sum\limits_{l = 1}^L {\sum\limits_{k = 1}^K {{\rm{Tr}}\left( {\omega _{l,k}}{{{\bf{W}}_{l,k}}{\bf{U}}_{l,k}^{\rm{H}}{{{\bf{\bar H}}}_{l,l,k}}{{\bf{F}}_{l,k}}} \right)} }  - \sum\limits_{l = 1}^L {\sum\limits_{k = 1}^K {{\rm{Tr}}\left({\omega _{l,k}} {{{\bf{W}}_{l,k}}{\bf{F}}_{l,k}^{\rm{H}}{\bf{\bar H}}_{l,l,k}^{\rm{H}}{{\bf{U}}_{l,k}}} \right)} } }
\end{array}
\\
&\textrm{s.t.}\quad 0 \le {\theta _m} \le 2\pi, m = 1, \cdots ,M, \label{djjewr}
\end{align}
\end{subequations}
where ${\bf{F}}_l=\sum\nolimits_{k = 1}^K{{\bf{F}}_{l,k}}{\bf{F}}_{l,k}^{\rm{H}}$.

By using  ${\bf{\bar H}}_{l,n,m} = {\bf{H}}_{n,m}^{r}{\bm{\Phi}} {{\bf{G}}_{l}^{r}} + {\bf{H}}_{l,n,m}$, we have
\begin{equation}\label{adsqa}
\begin{array}{l}
{\omega _{n,m}}{{\bf{W}}_{n,m}}{\bf{U}}_{n,m}^{\rm{H}}{{{\bf{\bar H}}}_{l,n,m}}{{\bf{F}}_l}{\bf{\bar H}}_{l,n,m}^{\rm{H}}{{\bf{U}}_{n,m}}\\
 = {\omega _{n,m}}{{\bf{W}}_{n,m}}{\bf{U}}_{n,m}^{\rm{H}}{{\bf{H}}_{n,m}^{r}}{\bm{\Phi}} {{\bf{G}}_{l}^{r}}{{\bf{F}}_l}{\bf{G}}_{l}^{r\rm{H}} {{\bm{\Phi}} ^{\rm{H}}}{\bf{H}}_{n,m}^{r\rm{H}}{{\bf{U}}_{n,m}} + {\omega _{n,m}} {{\bf{W}}_{n,m}}{\bf{U}}_{n,m}^{\rm{H}}{{\bf{H}}_{l,n,m}}{{\bf{F}}_l}{\bf{G}}_{l}^{r\rm{H}}{{\bm{\Phi}} ^{\rm{H}}}{\bf{H}}_{n,m}^{r\rm{H}} {{\bf{U}}_{n,m}}\\
 \quad +{\omega _{n,m}} {{\bf{W}}_{n,m}}{\bf{U}}_{n,m}^{\rm{H}}{{\bf{H}}_{n,m}^{r}}{\bm{\Phi}} {{\bf{G}}_{l}^{r}}{{\bf{F}}_l}{\bf{H}}_{l,n,m}^{\rm{H}}{{\bf{U}}_{n,m}} + {\omega _{n,m}} {{\bf{W}}_{n,m}}{\bf{U}}_{n,m}^{\rm{H}}{{\bf{H}}_{l,n,m}}{{\bf{F}}_l}{\bf{H}}_{l,n,m}^{\rm{H}}{{\bf{U}}_{n,m}}
\end{array}
\end{equation}
and
\begin{equation}\label{dwgtrh}
 {\omega _{l,k}}{{\bf{W}}_{l,k}}{\bf{U}}_{l,k}^{\rm{H}}{{{\bf{\bar H}}}_{l,l,k}}{{\bf{F}}_{l,k}} = {\omega _{l,k}} {{\bf{W}}_{l,k}}{\bf{U}}_{l,k}^{\rm{H}}{{\bf{H}}_{l,k}^{r}}{\bm{\Phi}}{{\bf{G}}_{l}^{r}}{{\bf{F}}_{l,k}} + {\omega _{l,k}} {{\bf{W}}_{l,k}}{\bf{U}}_{l,k}^{\rm{H}}{{\bf{H}}_{l,l,k}}{{\bf{F}}_{l,k}}.
\end{equation}

By defining
${{\bf{B}}_{n,m}} \buildrel \Delta \over =  {\omega _{n,m}}{\bf{H}}_{n,m}^{r \rm{H}}{{\bf{U}}_{n,m}}{{\bf{W}}_{n,m}}{\bf{U}}_{n,m}^{\rm{H}}{{\bf{H}}_{n,m}^{r}}$, ${{\bf{C}}_l} \buildrel \Delta \over =  {{\bf{G}}_{l}^{r}}{{\bf{F}}_l}{\bf{G}}_{l}^{r\rm{H}} $ and
\begin{equation*}
  {{\bf{D}}_{l,n,m}} \buildrel \Delta \over = {\omega _{n,m}}{{\bf{G}}_{l}^{r}}{\bf{F}}_l^{\rm{H}}{\bf{H}}_{l,n,m}^{\rm{H}}{{\bf{U}}_{n,m}}{{\bf{W}}_{n,m}}{\bf{U}}_{n,m}^{\rm{H}}{{\bf{H}}_{n,m}^{r}},
\end{equation*}
from (\ref{adsqa}) we have
\begin{equation}\label{wdfwfgr}
  \begin{array}{l}
{\rm{Tr}}\left({\omega _{n,m}} {{{\bf{W}}_{n,m}}{\bf{U}}_{n,m}^{\rm{H}}{{{\bf{\bar H}}}_{l,n,m}}{{\bf{F}}_l}{\bf{\bar H}}_{l,n,m}^{\rm{H}}{{\bf{U}}_{n,m}}} \right)\\
 = {\rm{Tr}}\left( {{{\bm{\Phi}} ^{\rm{H}}}{{\bf{B}}_{n,m}}{\bm{\Phi}} {{\bf{C}}_{l}}} \right) + {\rm{Tr}}\left( {{{\bm{\Phi}} ^{\rm{H}}}{{\bf{D}}_{l,n,m}^{\rm{H}}}} \right) + {\rm{Tr}}\left( {{\bm{\Phi}} {\bf{D}}_{l,n,m} } \right) + {\rm{const}_1},
\end{array}
\end{equation}
where ${\rm{const}_1}$ is a constant term that does not depend on ${\bm{\Phi}}$.

Similarly, by defining  ${{\bf{T}}_{l,k}} \buildrel \Delta \over = {\omega _{l,k}}{{\bf{G}}_{l}^{r}}{{\bf{F}}_{l,k}}{{\bf{W}}_{l,k}}{\bf{U}}_{l,k}^{\rm{H}}{{\bf{H}}_{l,k}^{r}}$, from (\ref{dwgtrh}) we have
\begin{equation}\label{edwf}
 {\rm{Tr}}\left( {{\omega _{l,k}}{{\bf{W}}_{l,k}}{\bf{U}}_{l,k}^{\rm{H}}{{{\bf{\bar H}}}_{l,l,k}}{{\bf{F}}_{l,k}}} \right) = {\rm{Tr}}\left({\bm{\Phi}} {{\bf{T}}_{l,k}}\right) + {\rm{const}}_2,
\end{equation}
where ${\rm{const}}_2$ is a constant term that is independent of ${\bm{\Phi}}$.

By substituting (\ref{wdfwfgr}) and (\ref{edwf}) into the OF of Problem (\ref{appsdsworig}) and ignoring the constant terms,  we have
 \begin{subequations}\label{appjswifhorig}
\begin{align}
&{\mathop {\min }\limits_{\bm{\theta}}  \quad {\rm{Tr}}\left( {{{\bm{\Phi}} ^{\rm{H}}}{\bf{B}}{\bm{\Phi}} {\bf{C}}} \right) + {\rm{Tr}}\left( {{{\bm{\Phi}} ^{\rm{H}}}{{\bf{V}}^{\rm{H}}}} \right) + {\rm{Tr}}\left( {{\bm{\Phi}} {\bf{V}}} \right)}
\\
&\textrm{s.t.}\quad 0 \le {\theta _m} \le 2\pi, m = 1, \cdots ,M, \label{djhuhur}
\end{align}
\end{subequations}
where ${\bf{B}}$, ${\bf{C}}$ and ${\bf{V}}$ are respectively given by
 \begin{equation}\label{zascdce}
   {\bf{B}} = \sum\limits_{n = 1}^L {\sum\limits_{m = 1}^K {{{\bf{B}}_{n,m}}} } ,{\bf{C}} = \sum\limits_{l = 1}^L {{{\bf{C}}_l}} ,{\bf{V}} = \sum\limits_{l = 1}^L {\sum\limits_{n = 1}^L {\sum\limits_{m = 1}^K {{{\bf{D}}_{l,n,m}}} } }  - \sum\limits_{l = 1}^L {\sum\limits_{k = 1}^K {{{\bf{T}}_{l,k}}} }.
 \end{equation}

Upon denoting the collection of diagonal elements of ${\bm{\Phi}}$ by ${\bm{\phi}} \buildrel \Delta \over = {\left[ {{e^{j{\theta _1}}}, \cdots ,{e^{j{\theta _m}}}, \cdots ,{e^{j{\theta _M}}}} \right]^{\rm{T}}}$ and using the matrix identity of \cite[Eq. (1.10.6)]{zhang2017matrix}, we arrive at
\begin{equation}\label{saddewde}
  {\rm{Tr}}\left( {{{\bm{\Phi}} ^{\rm{H}}}{\bf{B}}{\bm{\Phi}} {\bf{C}}} \right) = {{\bm{\phi}} ^{\rm{H}}}\left( {{\bf{B}} \odot {{\bf{C}}^{\rm{T}}}} \right){\bm{\phi}}.
\end{equation}
Let ${\bf{v}}$ be the collection of diagonal elements of matrix ${\bf{V}}$, given by ${\bf{v}} = {\left[ {{{\left[ {\bf{V}} \right]}_{1,1}}, \cdots ,{{\left[ {\bf{V}} \right]}_{M,M}}} \right]^{\rm{T}}}$. Then, we have
\begin{equation}\label{sdewf}
  {\rm{Tr}}\left( {{\bm{\Phi}} {\bf{V}}} \right)={\bm{\phi}}^{\rm{T}}{\bf{v}}, {\rm{Tr}}\left( {{{\bm{\Phi}} ^{\rm{H}}}{{\bf{V}}^{\rm{H}}}} \right) = {{\bf{v}}^{\rm{H}}}{{\bm{\phi}} ^*}.
\end{equation}
Hence, Problem (\ref{appjswifhorig}) can be rewritten as
 \begin{subequations}\label{appjhorig}
\begin{align}
&{\mathop {\min }\limits_{\bm{\theta}}  \quad {{\bm{\phi}} ^{\rm{H}}}{\bm{\Xi} }{\bm{\phi}} + {\bm{\phi}}^{\rm{T}}{\bf{v}} + {{\bf{v}}^{\rm{H}}}{{\bm{\phi}} ^*}}
\\
&\textrm{s.t.}\quad 0 \le {\theta _m} \le 2\pi, m = 1, \cdots ,M, \label{dshur}
\end{align}
\end{subequations}
where $\bm{\Xi}={{\bf{B}} \odot {{\bf{C}}^{\rm{T}}}} $. It can be readily verified that ${\bf{B}}$ and ${{\bf{C}}^{\rm{T}}}$ are semidefinite matrices. Then, according to Property (9) on Page 104 of \cite{zhang2017matrix}, the Hadamard product ${{\bf{B}} \odot {{\bf{C}}^{\rm{T}}}}$ (or equivalently $\bm{\Xi}$) is also a semidefinite matrix.

Recall that ${\phi _m} = {e^{j{\theta _m}}}, \forall m$, and that ${\bm{\phi}}={\left[ {{\phi _1}, \cdots ,{\phi _M}} \right]^{\rm{T}}}$. Then, Problem (\ref{appjhorig}) can be equivalently rewritten as
 \begin{subequations}\label{appjig}
\begin{align}
&{\mathop {\min }\limits_{\bm{\phi}}  \quad f({\bm{\phi}})\buildrel \Delta \over = {{\bm{\phi}} ^{\rm{H}}}{\bm{\Xi}}{\bm{\phi}} +  2{\rm{Re}}\left\{ {{{\bm{\phi}} ^{\rm{H}}}{{\bf{v}}^*}} \right\}}
\\
&\textrm{s.t.}\quad \left| {{\phi _m}} \right| = 1 , m = 1, \cdots ,M. \label{dshxsdceur}
\end{align}
\end{subequations}

Due to the unit modulus constraint in (\ref{dshxsdceur}), Problem (\ref{appjig}) is a non-convex optimization problem. In the following, we provide a pair of efficient algorithms for  solving this problem.

\subsubsection{Majorization-Minimization (MM) Algorithm}
 We  adopt the MM algorithm \cite{yansun} to solve Problem (\ref{appjig}), which was originally introduced in  \cite{marks1978general}. Then, this method has been widely in resource allocation for wireless communication networks \cite{ngo2013joint,venturino2009coordinated,zappone2015energy}.  The main idea is to solve a difficult problem by constructing a series of more tractable approximate subproblems. Specifically, let us denote the solution of the subproblem at the $t$th iteration by ${\bm{\phi}}^{t}$, and    the OF value of Problem (\ref{appjig}) at the $t$th iteration by $f({\bm{\phi}}^t)$. Then, at the $(t+1)$st iteration, we have to introduce an upper bound \footnote{Please note that  we consider the minimization problem here. } of the OF function based on the previous solution, which is denoted as $g({\bm{\phi}}|{\bm{\phi}}^t)$. We solve the approximate subproblem with the aid of the new OF $g({\bm{\phi}}|{\bm{\phi}}^t)$ at the $(t+1)$st iteration.  If the OF $g({\bm{\phi}}|{\bm{\phi}}^t)$ satisfies the following three conditions:
\begin{enumerate}
  \item $g({\bm{\phi}}^t|{\bm{\phi}}^t)=f({\bm{\phi}}^t)$,
  \item$\nabla_{\bm{\phi}}g({\bm{\phi}}|{\bm{\phi}}^t)|_{{\bm{\phi}}={\bm{\phi}}^t}=\nabla_{\bm{\phi}}f({\bm{\phi}}^t) |_{{\bm{\phi}}={\bm{\phi}}^t}$,
  \item $g({\bm{\phi}}|{\bm{\phi}}^t)\ge f({\bm{\phi}})$,
\end{enumerate}
then the sequence of the solutions obtained in each iteration will result in a monotonically decreasing OF $\{f({\bm{\phi}}^t), t=1,2, \cdots\}$ and finally converge. The converged solution  satisfies the Karush-Kuhn-Tucker (KKT) optimality conditions of Problem (\ref{appjig}) \cite{cunhuapanwcl}. The first two conditions represent that the OF $g({\bm{\phi}}|{\bm{\phi}}^t)$ introduced   and its first-order gradient should be the same as the original OF and its first-order gradient at point ${\bm{\phi}}^t$. The third condition means that the OF $g({\bm{\phi}}|{\bm{\phi}}^t)$ constructed  should represent the upper bound of the original OF. To make this algorithm work, the most important task is to find the OF $g({\bm{\phi}}|{\bm{\phi}}^t)$, which should satisfy these three conditions and should be much more tractable than $f({\bm{\phi}})$.

To this end, we first introduce the following lemma proposed in \cite{jiansong}.

\itshape \textbf{Lemma 1:}  \upshape  For any given solution ${\bm{\phi}}^t$ at the $t$th iteration and for any feasible ${\bm{\phi}}$, we have
\begin{equation}\label{sxdwef}
  {{\bm{\phi}} ^{\rm{H}}}{\bm{\Xi}}{\bm{\phi}}  \le {{\bm{\phi}} ^{\rm{H}}}{\bf{X}}{\bm{\phi}}  - 2{\mathop{\rm Re}\nolimits} \left\{ {{{\bm{\phi}} ^{\rm{H}}}\left( {{\bf{X}} - {\bm{\Xi}}} \right){{\bm{\phi}} ^t}} \right\} + {\left( {{{\bm{\phi}} ^t}} \right)^{\rm{H}}}\left( {{\bf{X}} - {\bm{\Xi}}} \right){{\bm{\phi}}^t} \buildrel \Delta \over = y({\bm{\phi}} |{{\bm{\phi}}^t}),
\end{equation}
where ${\bf{X}} = {\lambda _{\rm{\max} }}{{\bf{I}}_M}$ and ${\lambda _{\rm{\max} }}$ is the maximum eigenvalue of ${\bm{\Xi}}$.  \hfill $\Box$

Upon constructing the surrogate OF $g({\bm{\phi}}|{\bm{\phi}}^t)$ as follows:
\begin{equation}\label{sdwxs}
 g({\bm{\phi}}|{\bm{\phi}}^t)=y({\bm{\phi}} |{{\bm{\phi}}^t})+ 2{\rm{Re}}\left\{ {{{\bm{\phi}} ^{\rm{H}}}{{\bf{v}}^*}} \right\},
\end{equation}
where $y({\bm{\phi}} |{{\bm{\phi}}^t})$ is defined in (\ref{sxdwef}), it can be readily verified that $g({\bm{\phi}}|{\bm{\phi}}^t)$ given in (\ref{sdwxs}) satisfies the three conditions. Additionally, the OF $g({\bm{\phi}}|{\bm{\phi}}^t)$ is more tractable  than the original OF $f({\bm{\phi}})$.   Specifically, the subproblem to be solved at the $t$th iteration   is given by
 \begin{subequations}\label{apsasdcjig}
\begin{align}
&{\mathop {\min }\limits_{\bm{\phi}}  \quad g({\bm{\phi}}|{\bm{\phi}}^t)}
\\
&\textrm{s.t.}\quad \left| {{\phi _m}} \right| = 1 , m = 1, \cdots ,M. \label{dshxceur}
\end{align}
\end{subequations}
Since ${\bm{\phi}} ^{\rm{H}}{\bm{\phi}}=M$, we have  ${{\bm{\phi}} ^{\rm{H}}}{\bf{X}}{\bm{\phi}}=M\lambda _{\rm{\max} }$, which is a constant. By removing the other constants, Problem (\ref{apsasdcjig}) can be rewritten as follows:
 \begin{subequations}\label{asasdcjig}
\begin{align}
&{\mathop {\max }\limits_{\bm{\phi}}  \quad 2{\mathop{\rm Re}\nolimits} \left\{ {{{\bm{\phi}} ^{\rm{H}}\bf{q}}^t } \right\}}
\\
&\textrm{s.t.}\quad \left| {{\phi _m}} \right| = 1 , m = 1, \cdots ,M, \label{aadshxceur}
\end{align}
\end{subequations}
where ${\bf{q}}^t=\left( {{\lambda _{\rm{\max} }}{{\bf{I}}_M} - {\bm{\Xi}}} \right){{\bm{\phi}} ^t}-{\bf{v}}^*.$ The optimal solution of Problem (\ref{asasdcjig}) is given by
\begin{equation}\label{dcsd}
{{\bm{\phi}}^{t+1}} = {e^{j\arg ({{\bf{q}}^t})}}.
\end{equation}

Based on the above discussions, we provide the details of the MM algorithm in Algorithm \ref{iterdada}. When the algorithm converges, we can obtain the optimal phase shift as  ${{\bm{\theta}}^\star}=\arg ({{\bm{q}}^t})$.

\begin{algorithm}
\caption{MM Algorithm}\label{iterdada}
\begin{algorithmic}[1]
\STATE Initial the iteration number $t=1$, the accuracy $\varepsilon$. Input the feasible solution ${\bm{\phi}} ^0$. Calculate the value of the objective function in Problem (\ref{appjig}) as $f({{\bm{\phi}} ^1})$;
\STATE Calculate ${\bf{q}}^t=\left( {{\lambda _{\rm{\max} }}{{\bf{I}}_M} - {\bm{\Xi}}} \right){{\bm{\phi}} ^t}-{\bf{v}}^*$;
\STATE Update ${{\bm{\phi}}^{t+1}} $ in (\ref{dcsd});
 \STATE Calculate the objective function  $f({{\bm{\phi}} ^{t+1}})$, if ${{\left| {f({\bm{\phi} ^{t+1}}) - f({\bm{\phi} ^{t}})} \right|} \mathord{\left/
 {\vphantom {{\left| {f({\bm{\phi} ^{t+1}}) - f({\bm{\phi} ^{t }})} \right|} {f({\bm{\phi} ^{t+1}})}}} \right.
 \kern-\nulldelimiterspace} {f({\bm{\phi} ^{t+1}})}} \le \varepsilon $ holds, terminate; Otherwise, set $t \leftarrow t + 1$ and go to step 2.
\end{algorithmic}
\end{algorithm}

\subsubsection{Complex Circle Manifold (CCM) Method}
In this subsection, we adopt the CCM method proposed in  \cite{alhujaili2019transmit} for directly solving Problem (\ref{appjig}). We first transform Problem (\ref{appjig})  into the following equivalent  problem
 \begin{subequations}\label{appdbhig}
\begin{align}
&{\mathop {\min }\limits_{\bm{\phi}}  \quad \bar f({\bm{\phi}}) \buildrel \Delta \over =  {{\bm{\phi}} ^{\rm{H}}}(\bm{\Xi}+\alpha {{\bf{I}}_M}){\bm{\phi}} +  2{\rm{Re}}\left\{ {{{\bm{\phi}} ^{\rm{H}}}{{\bf{v}}^*}} \right\}}
\\
&\textrm{s.t.}\quad \left| {{\phi _m}} \right| = 1 , m = 1, \cdots ,M, \label{dsdceur}
\end{align}
\end{subequations}
where $\alpha>0$ is a positive constant parameter, the value of which will be given in Theorem 1. Problem  (\ref{appjig}) is equivalent to Problem (\ref{appdbhig}), since we have $\alpha{\bm{\phi}} ^{\rm{H}}{\bm{\phi}}=\alpha M$. The parameter $\alpha$ can control the convergence of the CCM method, which will be discussed in Theorem 1.

The search space in Problem (\ref{appdbhig}) can be regarded as the product of $M$ complex circles\footnote{Each complex circle is given by ${\cal S} \buildrel \Delta \over = \left\{ {x \in \mathbb{C}:{x^*}x = {\mathop{\rm Re}\nolimits} {{\left\{ x \right\}}^2} + {\mathop{\rm Im}\nolimits} {{\left\{ x \right\}}^2} = 1} \right\}$, which is a sub-manifold of $\mathbb{C}$ \cite{alhujaili2019transmit}.}, which is a sub-manifold of $\mathbb{C}^{M}$ given by
\begin{equation}\label{johoii}
  {{\cal S}^M} \buildrel \Delta \over = \left\{ { {\bf{x}} \in {\mathbb{C}^M}:\left| {{x_l}} \right| = 1,l = 1,2, \cdots ,M} \right\},
\end{equation}
where $x_l$ is the $l$th element of vector ${\bf{x}}$.

\begin{figure}
\centering
\includegraphics[width=3.2in]{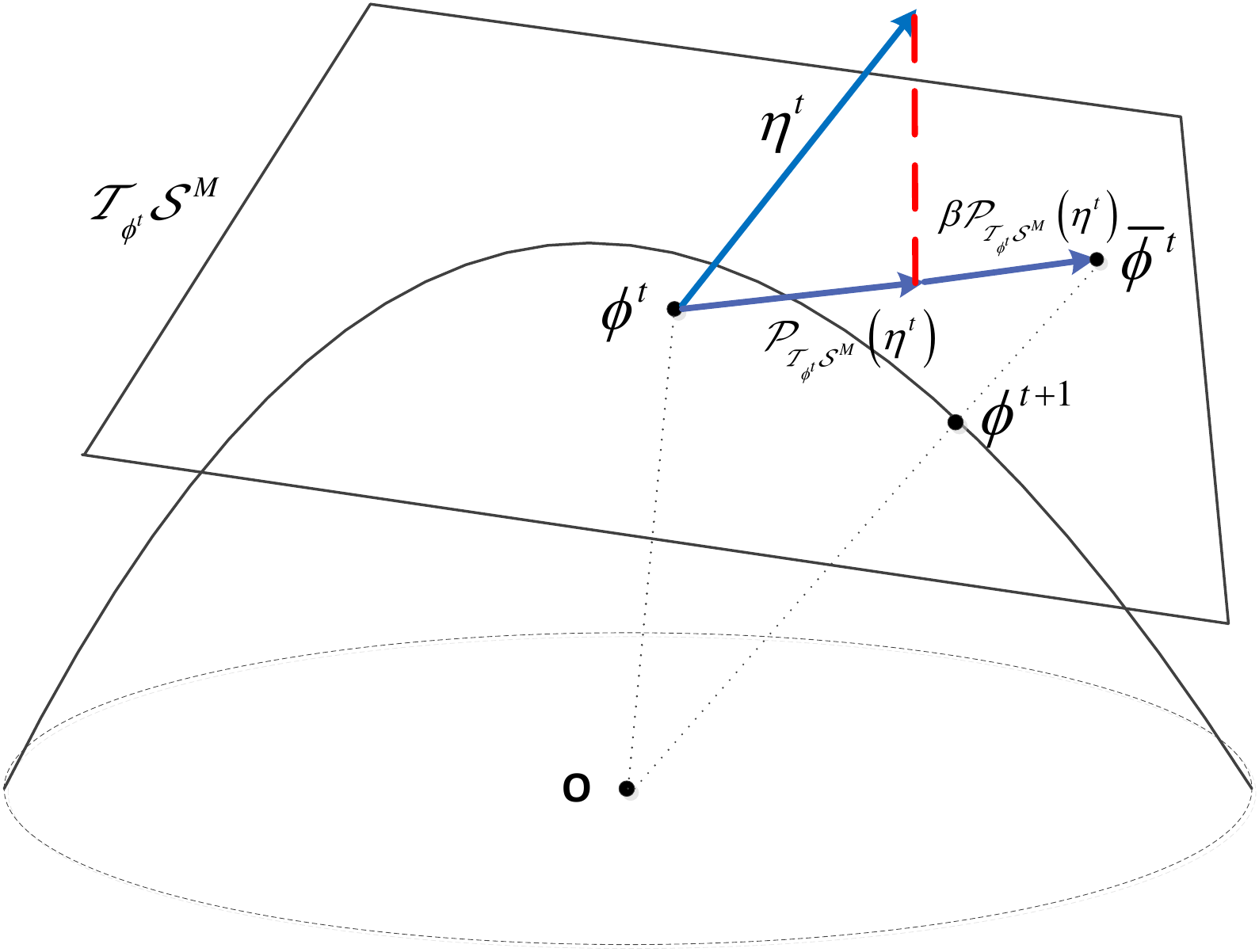}\vspace{-0.5cm}
\caption{Geometric interpretation of the CCM algorithm. }
\label{fig2}
\end{figure}

The main idea of the CCM algorithm is to derive a gradient descent algorithm based on the manifold space defined in  (\ref{johoii}), which is similar to the concept of the gradient descent technique developed for the conventional optimization over the Euclidean space. The main steps of the CCM algorithm is composed of four main steps in each iteration $t$:

1) \emph{Gradient in Euclidean Space:} We first have to find the search direction and the most common search direction for a minimization problem is to move in the direction opposite to the gradient of $\bar f({\bm{\phi}^t})$, which is given by
\begin{equation}\label{xsfrar}
 {\bm{\eta} ^t} =  - {\nabla _{\bm{\phi}} }\bar f({{\bm{\phi}} ^t}) =  - 2(\bm{\Xi}  + \alpha {{\bf{I}}_M}){{\bm{\phi}} ^t} - 2{{\bf{v}}^*}.
\end{equation}

2) \emph{Riemannian gradients:} Since we optimize over the manifold space, we have to find the Riemannian gradient \cite{yu2016alternating}.  The Riemannian gradient of $\bar f({\bm{\phi}^t})$ at the current point ${\bm{\phi}^t}\in {\cal S}^M$ is in the tangent space ${\cal T}_{{\bm{\phi}^t}}{\cal S}^M$ \footnote{The tangent space of ${\cal S}$ at point $z_m$ is defined as  ${\cal T}_{z_m}{\cal S}=\{x\in \mathbb{C}: {\rm{Re}}\{x^*z_m\}=0\}$. Then, the tangent space ${\cal T}_{{\bf{z}}}{\cal S}^M$ is the product of these $M$  tangent space ${\cal T}_{z_m}{\cal S}$ given by ${\cal T}_{{\bf{z}}}{\cal S}^M={\cal T}_{{z}_1}{\cal S}\times{\cal T}_{{z}_2}{\cal S}\cdots \times {\cal T}_{{z}_M}{\cal S}$.}. Specifically, the Riemannian gradient of $\bar f({\bm{\phi}^t})$ at  ${\bm{\phi}^t}$ can be obtained by projecting the search direction ${\bm{\eta} ^t}$ in the Euclidean space onto  ${\cal T}_{{\bm{\phi}^t}}{\cal S}^M$ by using the projection operator, which can be calculated as follows \cite{yu2016alternating}:
\begin{equation}\label{dftrhyu}
 {\bf{P}}_{{\cal T}_{{\bm{\phi}^t}}{\cal S}^M}({\bm{\eta} ^t})={\bm{\eta} ^t}-{\rm{Re}}\{\bm{\eta} ^{t*}\odot {\bm{\phi}^t}\}\odot{\bm{\phi}^t}.
\end{equation}

3) \emph{Update over the tangent space:} Update the current point ${\bm{\phi}^t}$ on the tangent space ${{\cal T}_{{\bm{\phi}^t}}{\cal S}^M}$:
\begin{equation}\label{sfrtg}
 \bar {\bm{\phi}^t}={\bm{\phi}^t}+\beta{\bf{P}}_{{\cal T}_{{\bm{\phi}^t}}{\cal S}^M}({\bm{\eta} ^t}),
\end{equation}
where $\beta$ is a constant step size that will be discussed in Theorem 1.

4) \emph{Retraction operator:} In general, the  $\bar {\bm{\phi}^t}$ obtained is not in ${\cal S}^M$, i.e.
we have $\bar {\bm{\phi}^t} \notin {\cal S}^M$. Hence, it has to be mapped into the manifold ${\cal S}^M$ by using the
retraction operator\footnote{The retraction operator normalizes each element of $\bar {\bm\phi}^t $ to be unit.} as follows
\begin{equation}\label{sdwegt}
  {\bm{\phi}^{t+1}}=\bar {\bm\phi}^t  \odot \frac{1}{{\left| \bar{ \bm\phi}^t  \right|}}.
\end{equation}

Note that both ${\bm{\phi}^{t+1}}$ and ${\bm{\phi}^{t}}$ belongs to ${{\cal S}^M}$, which satisfies the unit constant modulus constraints. The details of the CCM algorithm are presented in Algorithm \ref{itszsrda}. The CCM algorithm is also illustrated  geometrically  in Fig.~\ref{fig2}.

\begin{algorithm}
\caption{CCM Algorithm}\label{itszsrda}
\begin{algorithmic}[1]
\STATE Initial the iteration number $t=1$, the accuracy $\varepsilon$. Input the feasible solution ${\bm{\phi}} ^1$. Calculate the value of the objective function in Problem (\ref{appdbhig}) as $\bar f({{\bm{\phi}} ^1})$;
\STATE Calculate the Euclidean gradient $ {\bm{\eta} ^t}$ in  (\ref{xsfrar});
\STATE Calculate the Riemannian gradient ${\bf{P}}_{{\cal T}_{{\bm{\phi}^t}}{\cal S}^M}({\bm{\eta} ^t})$ in (\ref{dftrhyu});
\STATE Update over the tangent space  according to (\ref{sfrtg});
\STATE Update ${\bm{\phi}^{t+1}}$ by retracting $\bar {\bm{\phi}^t}$ to the complex circle manifold $ {{\cal S}^M}$ according to (\ref{sdwegt});
 \STATE Calculate the objective function  $\bar f({{\bm{\phi}} ^{t+1}})$, if ${{\left| {\bar f({\bm{\phi} ^{t+1}}) - \bar f({\bm{\phi} ^{t }})} \right|} \mathord{\left/
 {\vphantom {{\left| {\bar f({\bm{\phi} ^{t+1}}) - \bar f({\bm{\phi} ^{t}})} \right|} {\bar f({\bm{\phi} ^{t+1}})}}} \right.
 \kern-\nulldelimiterspace} {\bar f({\bm{\phi} ^{t+1}})}} \le \varepsilon $ holds, terminate; Otherwise, set $t \leftarrow t + 1$ and go to step 2.
\end{algorithmic}
\end{algorithm}

The following theorem provides  guidance for the choices of parameters $\alpha$ and $\beta$ to guarantee the convergence of the CCM algorithm.

\itshape \textbf{Theorem 1 \cite{alhujaili2019transmit}:}  \upshape  Let $\lambda _{\bm{\Xi}}$ and $\lambda _{\bm{\Xi}+\alpha {{\bf{I}}_M}}$ be the largest eigenvalue of matrices $\bm{\Xi}$ and $\bm{\Xi}+\alpha {{\bf{I}}_M}$, respectively. If $\alpha$ and $\beta$ are chosen to satisfy the following conditions,
\begin{equation}\label{jrisstru}
  \alpha  \ge \frac{M}{8}{\lambda _{\bm{\Xi}} } + {\left\| {\bf{v}} \right\|_2},0 < \beta  < \frac{1}{{{\lambda _{\bm{\Xi}  + \alpha {\bf{I}}}}}},
\end{equation}
then the CCM algorithm generates a non-increasing sequence $\{\bar f({\bm{\phi} ^t}), t=1,2,\cdots \}$, and finally converges to a finite value.  \hfill $\Box$

\subsubsection{Complexity Analysis}

In this part, we analyze the complexity of both proposed   methods in solving Problem (\ref{appjig}).

Let us now analyze the complexity of the MM algorithm. At the beginning of the MM algorithm, we have to calculate ${\lambda _{\rm{\max} }}$, i.e.  the maximum eigenvalue of ${\bm{\Xi}}$. The associated complexity is given by ${\cal O}(M^3)$. For each iteration of the MM algorithm, the main complexity lies in the calculation of ${\bf{q}}^t$ in Step 2, the complexity of which is ${\cal O}(M^2)$. Let us denote  the number of iterations required for the MM algorithm to converge by $T_{MM}$. Then, the total complexity of the MM algorithm is given by $C_{\rm{MM}}={\cal O}(M^3+T_{MM}M^2)$.

We then analyze the complexity of the CCM algorithm. At the start of the CCM algorithm, we have to find the range of
$\alpha $ and $\beta$ to guarantee the convergence of the CCM algorithm, which relies on calculating the largest eigenvalue of the matrices $\bm{\Xi}$ ($\lambda _{\bm{\Xi}}$), as shown in Theorem 1. Its complexity order is given by ${\cal O}(M^3)$. For each iteration of the CCM algorithm, the complexity  mainly depends on the calculation of the Euclidean gradient ${\bm{\eta} ^t} $, which is given by ${\cal O}(M^2)$. Let us denote the total number of iterations required by the CCM algorithm to converge by $T_{CCM}$. Then, the total complexity of the CCM algorithm is given by $C_{\rm{CCM}}={\cal O}(M^3+T_{CCM}M^2)$.

The complexity of these algorithms is summarized in Table \ref{tabzero}. It can be observed that the complexity mainly depends on the number of iterations required for convergence. The simulation results of Section \ref{simlresult} will compare their convergence speed.

\begin{table}[!t]
\renewcommand{\arraystretch}{1.1}
\caption{Computational Complexity Comparison for Two Different Algorithms to Find the Phase Shifts}
\label{tabzero}
\centering
\begin{tabular}{|c|c|c|}
\hline
Algorithms    & MM Alg. & CCM Alg.  \\
\hline
Complexity  & ${\cal O}(M^3+T_{MM}M^2)$ & ${\cal O}(M^3+T_{CCM}M^2)$\\
\hline
\end{tabular}
\end{table}

\subsection{Overall Algorithm to Solve Problem (\ref{appstaoneorig})}

Based on the above analysis, we provide the detailed description of the BCD algorithm conceived for solving Problem (\ref{appstaoneorig}) in Algorithm \ref{bcd}.  In Step 5, we have to apply two algorithms for solving Problem (\ref{appjig}) to  find the phase shifts ${\bm{\theta}}^{(n+1)}$. Both the MM algorithm and the CCM algorithm can guarantee to yield a monotonically decreasing OF value of Problem (\ref{appjig}) compared to the previous phase solution, i.e., $f({\bm{\phi}}^{(n+1)})<f({\bm{\phi}}^{(n)})$.  It can be readily verified that the OF value of Problem (\ref{appstneorig}) monotonically increases in each step of
Algorithm \ref{bcd}. Additionally, due to the power constraints, the OF value has an upper bound. Hence, Algorithm \ref{bcd} is guaranteed to converge.

\begin{algorithm}
\caption{Block Coordinate Descent Algorithm}\label{bcd}
\begin{algorithmic}[1]
\STATE Initialize iterative number $n=1$, maximum number of iterations $n_{\rm{max}}$,  feasible ${\bf{F}}^{(1)}$,
${\bm{\theta}}^{(1)}$, error tolerance $\varepsilon$, calculate the OF value of Problem (\ref{appstaoneorig}), denoted as ${\rm{Obj(}}{{\bf{F}}^{(1)}},{\bm{\theta}}^{(1)}{\rm{)}}$;
\STATE Given ${\bf{F}}^{(n)}$ and ${\bm{\theta}}^{(n)}$, calculate the optimal decoding matrices ${\bf{U}}^{ (n)}$ in (\ref{fgrtgtyh});
\STATE Given ${\bf{F}}^{(n)}$, ${\bf{U}}^{ (n)}$ and ${\bm{\theta}}^{(n)}$, calculate the optimal auxiliary matrices ${\bf{W}}^{(n)}$;
\STATE Given ${\bf{U}}^{ (n)}$, ${\bf{W}}^{(n)}$ and ${\bm{\theta}}^{(n)}$, calculate the optimal  precoding matrices ${\bf{F}}^{(n+1)}$ by solving Problem (\ref{appssxsorig}) with the Lagrangian multiplier method in Subsection \ref{kodsijcosakpdc};
 \STATE Given ${\bf{U}}^{ (n)}$, ${\bf{W}}^{(n)}$ and ${\bf{F}}^{(n+1)}$, calculate the optimal ${\bm{\theta}}^{(n+1)}$ by solving Problem (\ref{appjig}) with the algorithms developed in Subsection \ref{hwdi};
 \STATE If $n\geq n_{\rm{max}}$ or ${{\left| {{\rm{Obj}}({{\bf{F}}^{(n + 1)}},{{\bm{\theta}} ^{(n + 1)}}) - {\rm{Obj}}({{\bf{F}}^{(n)}},{{\bm{\theta}}^{(n)}})} \right|} \mathord{\left/
 {\vphantom {{\left| {{\rm{Obj}}({{\bf{F}}^{(n + 1)}},{{\bm{\theta}}^{(n + 1)}}) - {\rm{Obj}}({{\bf{F}}^{(n)}},{{\bm{\theta}} ^{(n)}})} \right|} {{\rm{Obj}}({{\bf{F}}^{(n + 1)}},{{\bm{\theta}} ^{(n + 1)}})}}} \right.
 \kern-\nulldelimiterspace} {{\rm{Obj}}({{\bf{F}}^{(n + 1)}},{{\bm{\theta}} ^{(n + 1)}})}} < \varepsilon$, terminate.  Otherwise, set $n \leftarrow n + 1$  and go to step 2.
\end{algorithmic}
\end{algorithm}

Let us now analyze the  complexity of the BCD algorithm. In Step 2, the complexity of computing the decoding matrices ${\bf{U}}^{ (n)}$ is ${\cal O}(LKN_r^3)$. In Step 3, the complexity of calculating the auxiliary matrices ${\bf{W}}^{(n)}$ is given by ${\cal O}(LKd^3)$. In Step 4, we have to calculate the TPC matrices ${\bf{F}}^{(n+1)}$. The detailed analysis is provided as follows. For any pair of complex matrices ${\bf{X}} \in {{\mathbb{C}}^{m \times n}},{\bf{Y}} \in {{\mathbb{C}}^{n \times p}}$, the complexity of computing ${\bf{XY}}$ is  ${\cal O}\left( {mnp} \right)$ \cite{boyd2004convex}. We assume that $N_t>N_r>d$.  Hence, the complexity of computing the matrices $\{{{{\bf{A}}_{l,k}}},\forall l,k\}$ in (\ref{wsdwe}) is given by  ${\cal O}(LKN_t^2d)$. The complexity of calculating ${{\bf{F}}_{l,k}}$ in (\ref{dwfohu}) is given by ${\cal O}(LKN_t^3)$. The SVD decomposition of $\{{\bf{A}}_l, \forall l\}$ is given by ${\cal O}(LN_t^3)$. The complexity of calculating $\{{\bf{Z}}_l\}$ is given by ${\cal O}(L^2N_t^2N_r)$. The complexity of evaluating the Lagrangian multipliers $\{\lambda _l,\forall l\}$ can be ignored. Hence, the overall complexity of calculating the TPC matrices ${\bf{F}}^{(n+1)}$ is given by ${\cal O}({\rm{max}}\{LKN_t^3,L^2N_t^2N_r\})$. The complexity of calculating the optimal ${\bm{\theta}}^{(n+1)}$ is given in Table \ref{tabzero}, while the complexity of each algorithm is denoted by $C_{i},i=\rm{ MM, CCM}$. Then, the overall complexity of the BCD algorithm is given by
 \begin{equation}\label{aefar}
   C_{\rm{BCD},i}={\cal O}({\rm{max}}\{LKN_t^3,L^2N_t^2N_r,C_{i}\}), i=\rm{ MM, CCM},
 \end{equation}
 where $C_{\rm{BCD},i}$ denotes the overall complexity of the BCD algorithm, when the phase shifts are obtained by using method $i$, $i=\rm{MM, CCM}$.
\section{ Extension to Other Scenarios }

\subsection{ Network MIMO }\label{joshgtro}

 In network MIMO, multiple BSs in different cells cooperate with each other and send the same data to each user. In this scenario, the antennas of all BSs form a giant antenna array  and jointly serve each user, where the inter-cell interference can be effectively mitigated \cite{cunhuamaga,cpan2017,cpanTWC2019}. It should be emphasized that compared to the model in Section \ref{system}, the data should be shared among multiple BSs, which incurs increased information exchange overhead.

 Let ${\bf{F}}_{i,l,k}$ be the precoding matrix of the $i$th BS for the $k$th  user in the $l$th cell, and ${{\bf{F}}_{l,k}} = {\left[ {{\bf{F}}_{i,l,k}^{\rm{H}},\forall i} \right]^{\rm{H}}} \in {\mathbb{C}^{L{N_t} \times d}}$ be the overall precoding matrix from all BSs to the user.
Define ${{\bf{G}}^r} = \left[ {{{\bf{G}}_{i}^{r}},\forall i} \right] \in {\mathbb{C}^{M \times L{N_t}}}$ be the overall channel from all the BSs to the IRS, and ${{\bf{H}}_{l,k}} = \left[ {{{\bf{H}}_{i,l,k}},\forall i} \right] \in {\mathbb{C}^{{N_r} \times L{N_t}}}$ the direct channel from all the BSs to the $k$th user in the $l$th cell. Let ${\bf{\bar H}}_{l,k} \buildrel \Delta \over =  {\bf{H}}_{r,l,k}{\bm{\Phi}} {{\bf{G}}^{r}} + {\bf{H}}_{l,k}$ be the equivalent channel spanning from all the BSs to the $k$th user in the $l$th cell.

 Then, the signal received at the $k$th user in the $l$th cell is given by
\begin{equation}\label{vfdtg}
 {{\bf{y}}_{l,k}} = {{{\bf{\bar H}}}_{l,k}}{{\bf{F}}_{l,k}}{{\bf{s}}_{l,k}} + \sum\limits_{m = 1,m \ne k}^K {{{{\bf{\bar H}}}_{l,k}}{{\bf{F}}_{l,m}}{{\bf{s}}_{l,m}}}  + \sum\limits_{i = 1,i \ne l}^L {\sum\limits_{m = 1}^K {{{{\bf{\bar H}}}_{l,k}}{{\bf{F}}_{i,m}}{{\bf{s}}_{i,m}}} }  + {{\bf{n}}_{l,k}}.
\end{equation}
The data rate of the $k$th user in the $l$th cell is given by
\begin{equation}\label{hdwewdreu}
{R_{l,k}}\left( {{\bf{F}},\bm{\theta} } \right) = {\log}\left| {{\bf{I}} + {{{\bf{\bar H}}}_{l,k}}{{\bf{F}}_{l,k}}{\bf{F}}_{l,k}^{\rm{H}}{\bf{\bar H}}_{l,k}^{\rm{H}}{\bf{J}}_{l,k}^{ - 1}} \right|,
\end{equation}
where ${\bf{J}}_{l,k}$ is given by
\begin{equation}\label{oswdwgth}
 {{\bf{J}}_{l,k}} = \sum\limits_{m = 1,m \ne k}^K {{{{\bf{\bar H}}}_{l,k}}{{\bf{F}}_{l,m}}{\bf{F}}_{l,m}^{\rm{H}}{\bf{\bar H}}_{l,k}^{\rm{H}}}  + \sum\limits_{i = 1,i \ne l}^L {\sum\limits_{m = 1}^K {{{{\bf{\bar H}}}_{l,k}}{{\bf{F}}_{i,m}}{\bf{F}}_{i,m}^{\rm{H}}{\bf{\bar H}}_{l,k}^{\rm{H}}} }  + {\sigma ^2}{\bf{I}}.
\end{equation}
The weighted sum rate problem is the same as in (\ref{appstaoneorig}), except that the power constraint for each BS is formulated as follows:
\begin{equation}\label{derfe}
\sum\limits_{l = 1}^L {\sum\limits_{k = 1}^K {\left\| {{{\bf{F}}_{i,l,k}}} \right\|_F^2} }  \le {P_{i,\max }},i = 1, \cdots ,L.
\end{equation}
The optimization problem formulated for the case of Network MIMO can be similarly solved by using the methods of Section \ref{algo}, details of which are omitted for simplicity.

\subsection{Multiple-IRS Scenario}

 Assume that the system has $A$ IRSs, each of which has $M$ reflection elements. The baseband channels spanning from  the $a$th IRS to the $k$th user in the $l$th cell, and the ones from the $i$th BS to the $a$th IRS are denoted by   ${\bf{H}}_{a,l,k}^r$ and ${\bf{G}}_{i,a}^r$, respectively. The diagonal phase-shifting matrix of the $a$th IRS is denoted by ${\bm{\Phi}_a}  = {\rm{diag}}\left\{ {{e^{j{\theta _{a,1}}}}, \cdots ,{e^{j{\theta _{a,m}}}}, \cdots ,{e^{j{\theta _{a,M}}}}} \right\}$. Then, the received signal  vector at the $k$th user in the $l$th cell is given by
\begin{equation}\label{uhwedwehu}
{{\bf{y}}_{l,k}} =    \sum\limits_{n = 1}^L {{{\bf{H}}_{n,l,k}}{{\bf{x}}_n}}    + \sum\limits_{n = 1}^L {\sum\limits_{a = 1}^A {{\bf{H}}_{a,l,k}^r{{\bm{\Phi}} _a}{\bf{G}}_{n,a}^r{{\bf{x}}_n}} }   + {{\bf{n}}_{l,k}},
\end{equation}
where ${\bf{H}}_{n,l,k}$, ${\bf{x}}_n$ and ${{\bf{n}}_{l,k}}$ are defined in Section \ref{system}.

 By defining ${\bf{H}}_{l,k}^r = \left[ {{\bf{H}}_{1,l,k}^r, \cdots ,{\bf{H}}_{A,l,k}^r} \right]$, ${\bm{\Phi}}= {\rm{diag}}\left\{ {{\bm{\Phi}}_1, \cdots,{\bm{\Phi}}_A} \right\}$ and ${\bf{G}}_n^r = {\left[ {{\bf{G}}_{n,1}^{r{\rm{H}}}, \cdots ,{\bf{G}}_{n,A}^{r{\rm{H}}}} \right]^{\rm{H}}}$, (\ref{uhwedwehu}) can be rewritten as
 \begin{equation}\label{uhdeweu}
{{\bf{y}}_{l,k}} =  \sum\limits_{n = 1}^L {{{\bf{H}}_{n,l,k}}{{\bf{x}}_n}}   +    \sum\limits_{n = 1}^L {{{\bf{H}}_{ l,k}^r}{\bm{\Phi}} {{\bf{G}}_{n}^r}} {{\bf{x}}_n}  + {{\bf{n}}_{l,k}},
\end{equation}
which is the same as (\ref{uhuhhu}). Hence, the derivations for the single-IRS scenario  are directly applicable.

\section{Simulation Results}\label{simlresult}

In this section,  simulation results are provided for validating the benefits of employing IRSs for improving WSR  of  multicell systems.  The large-scale path loss in dB is given by
\begin{equation}\label{shqidh}
  {\rm{PL = }}{{\rm{P}}{{\rm{L}}_0} - 10\alpha {{\log }_{10}}\left( {\frac{d}{{{d_0}}}} \right)},
\end{equation}
where ${\rm{P}}{{\rm{L}}_0}$ is the path-loss at the reference distance $d_0$, $d$ is the link distance, $\alpha$ is the path-loss exponent. In our simulations, we set ${\rm{P}}{{\rm{L}}_0}=-30$ dB and $d_0=1$ m. Due to extensive obstacles and scatterers, the path-loss exponent between the BS and the users is given by ${\alpha _{{\rm{BU}}}} = 3.75$.  The heights of BSs, IRSs, and users are assumed to be 30 m, 10 m, and 1.5 m, respectively.   By carefully choosing the location of the IRS, the IRS-aided link has a higher probability of experiencing nearly free-space path loss.  Then, we set the path-loss exponents of the BS-IRS link and of the IRS-user link to ${\alpha _{{\rm{BI}}}}={\alpha _{{\rm{IU}}}} \buildrel \Delta \over = {\alpha _{{\rm{IRS}}}}= 2.2 $. For the direct channel from the BSs to users, the small-scale fading is assumed to be Rayleigh fading due to extensive scatters. However, for the IRS-related channels, the small-scale fading is assumed to be Rician fading. In specific, the small-scale channel can be modeled as
\begin{equation}\label{fvgbhnjik}
  {{\bf{\tilde H}}} = \sqrt {\frac{{{\beta}}}{{{\beta} + 1}}} {\bf{\tilde H}}^{{\rm{LoS}}} + \sqrt {\frac{1}{{{\beta} + 1}}} {\bf{\tilde H}}^{{\rm{NLoS}}},
 \end{equation}
where  $\beta$ is the Rician factor, ${\bf{\tilde H}}^{{\rm{LoS}}}$ is the deterministic line of sight (LoS), and ${\bf{\tilde H}}^{{\rm{NLoS}}}$ is the non-LoS (NLoS) component that is  Rayleigh fading. The LoS component
 ${\bf{\tilde H}}^{{\rm{LoS}}}$ is given by  ${\bf{\tilde H}}^{{\rm{LoS}}} = {{\bf{a}}_{{D_r}}}\left( {\vartheta^{AoA}} \right){\bf{a}}_{D_t}^H\left( {\vartheta^{AoD}} \right)$, where ${{\bf{a}}_{{D_r}}}\left( {\vartheta^{AoA}} \right)$  is defined as
 \begin{equation}\label{tgthbty}
  {{\bf{a}}_{{D_r}}}\left( {\vartheta^{AoA}} \right) = {\left[ {1,{e^{j\frac{{2\pi d}}{\lambda }\sin \vartheta^{AoA}}}, \cdots ,{e^{j\frac{{2\pi d}}{\lambda }({D_r} - 1)\sin \vartheta^{AoA}}}} \right]^T}
 \end{equation}
 and
\begin{equation}\label{jpjookko}
  {{\bf{a}}_{D_t}}\left( {\vartheta^{AoD}} \right) = {\left[ {1,{e^{j\frac{{2\pi d}}{\lambda }\sin \vartheta ^{AoD}}}, \cdots ,{e^{j\frac{{2\pi d}}{\lambda }(D_t - 1)\sin \vartheta^{AoD}}}} \right]^T}.
\end{equation}
In (\ref{tgthbty}) and (\ref{jpjookko}), $D_r$ and $D_t$ are the number of antennas/elements at the receiver side and transmitter side, respectively, $d$ is the antenna separation distance, $\lambda$ is the wavelength, $\vartheta^{AoD}$ is the angle of departure and $\vartheta^{AoA}$ is the angle of arrival. It is assumed that $\vartheta^{AoD}$ and  $\vartheta^{AoA}$ are randomly distributed within $[0,2\pi ]$. For simplicity, we set $d/\lambda=1/2$.   Unless otherwise stated, we set the simulation parameters as follows: Channel bandwidth of 10 MHz,  noise power density of $-174$ dBm/Hz, number of transmit antennas of $N_t=4$, number of receive antennas of $N_r=2$, number of data streams of $d=2$, number of reflection elements of $M=50$, maximum BS power of $P_{l,\rm{max}}=1 \ {\rm{W}},\forall l$, Rician factor of $\beta=3$, error tolerance of $\varepsilon=10^{-6}$, and weighting factor of $\omega_{l,k}=1, \forall l,k$. The $x$ coordinate of the center point of the first circle is given by $x_u=280$ m, which means that the users are located at the edge of their corresponding cells. The following results are obtained by averaging over 200 independent channel generations. In Step 5 of the BCD algorithm, if the MM method is used, the BCD algorithm is denoted as BCD-MM. Similar definition holds for BCD-CCM. The step parameters $\alpha$ and $\beta $ in the CCM algorithm are  set based on Theorem 1.

\begin{figure}
\centering
\includegraphics[width=3.2in]{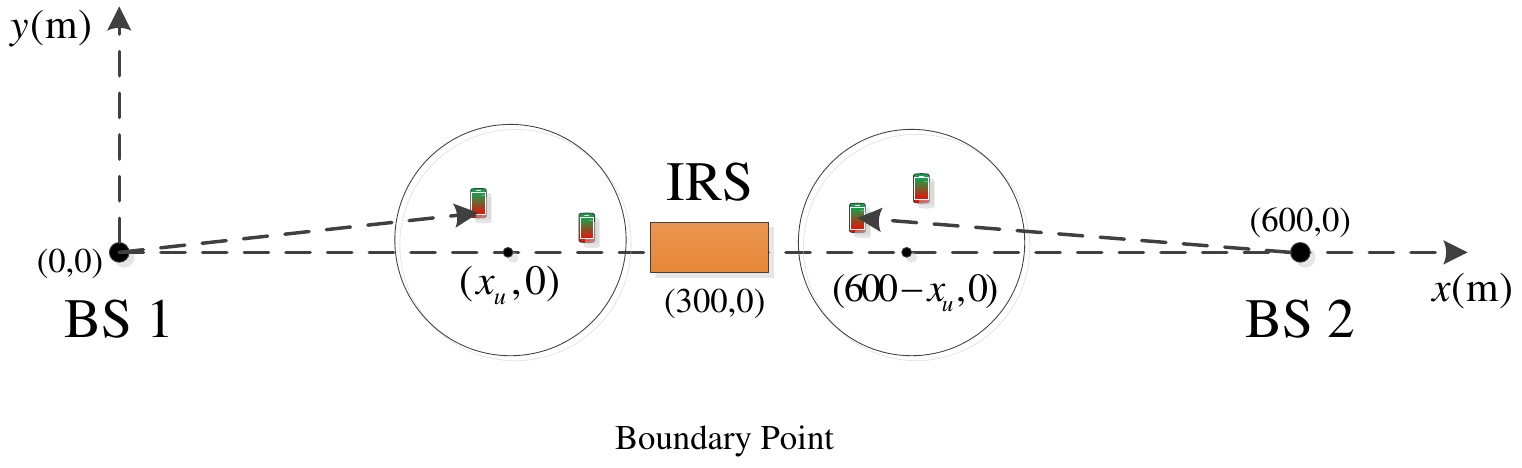}\vspace{-0.2cm}
\caption{The simulated two-cell IRS-aided MIMO communication scenario. }
\label{fig3}\vspace{-0.5cm}
\end{figure}

\subsection{Two-cell Scenario}
 In order to obtain more insights about the benefits of deploying IRS, we first consider a two-cell communication network with a single IRS  shown in Fig.~\ref{fig3}, in which there are two BSs located at $(0,0)$ and $(600,0)$\footnote{We only illustrate the horizontal plane of the system, where the height of various devices are not shown.}, respectively. By default, the IRS is deployed at the boundary point between two cells, the coordinate of which is $(300,0)$. Two users in the first cell are uniformly and randomly placed in a circle centered at $(x_u,0)$ with radius 20 m, while two users in the second cell are also uniformly and randomly distributed in a circle centered at $(600-x_u,0)$ with radius 20 m. Note that these two circles are symmetric w.r.t. the boundary point.
\subsubsection{Convergence Behaviour of BCD Algorithm}
\begin{figure}
\begin{minipage}[t]{0.495\linewidth}
\centering
\includegraphics[width=2.6in]{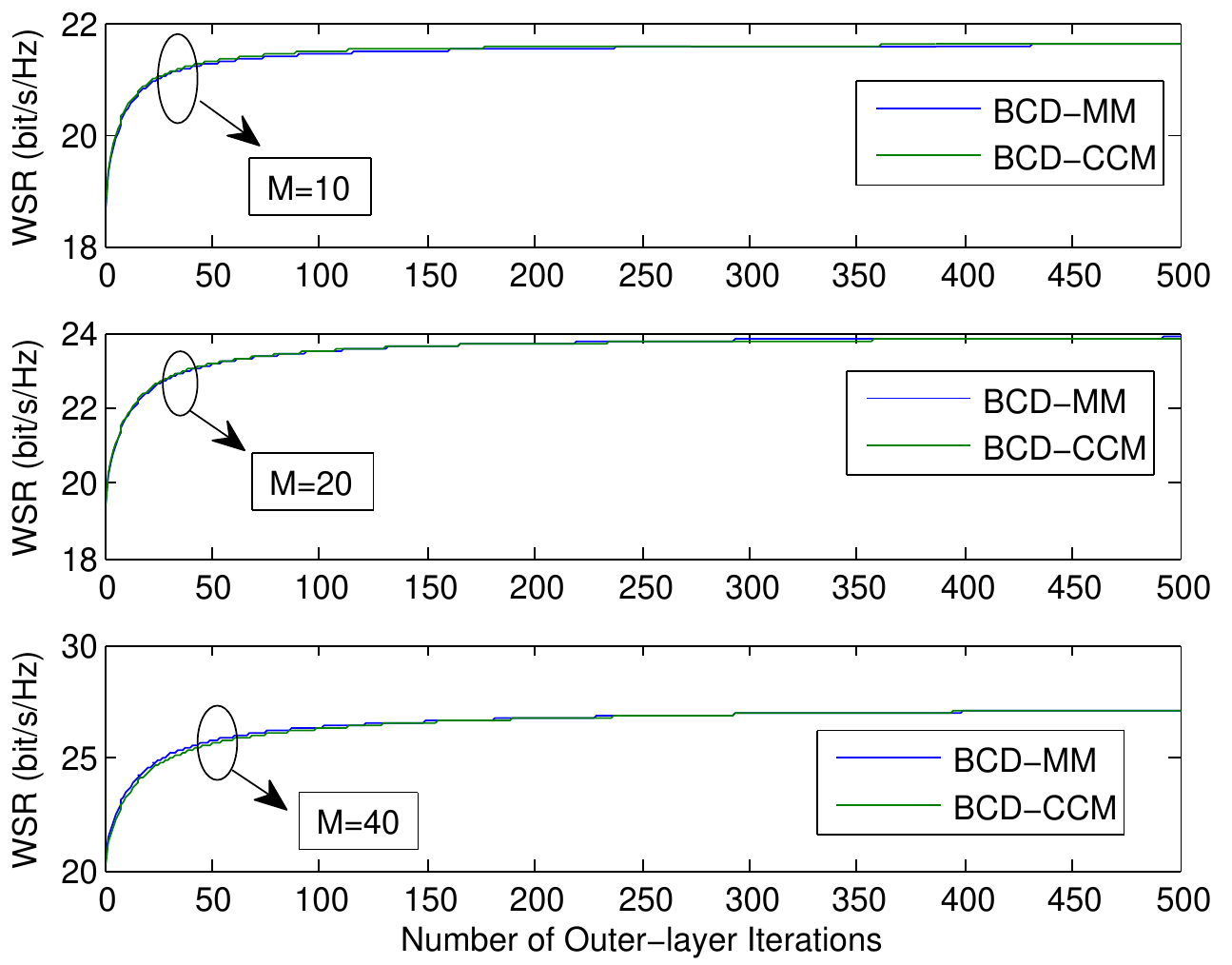}\vspace{-0.6cm}
\caption{Convergence behaviour of the BCD algorithm.}
\label{fig4}
\end{minipage}%
\hfill
\begin{minipage}[t]{0.495\linewidth}
\centering
\includegraphics[width=2.6in]{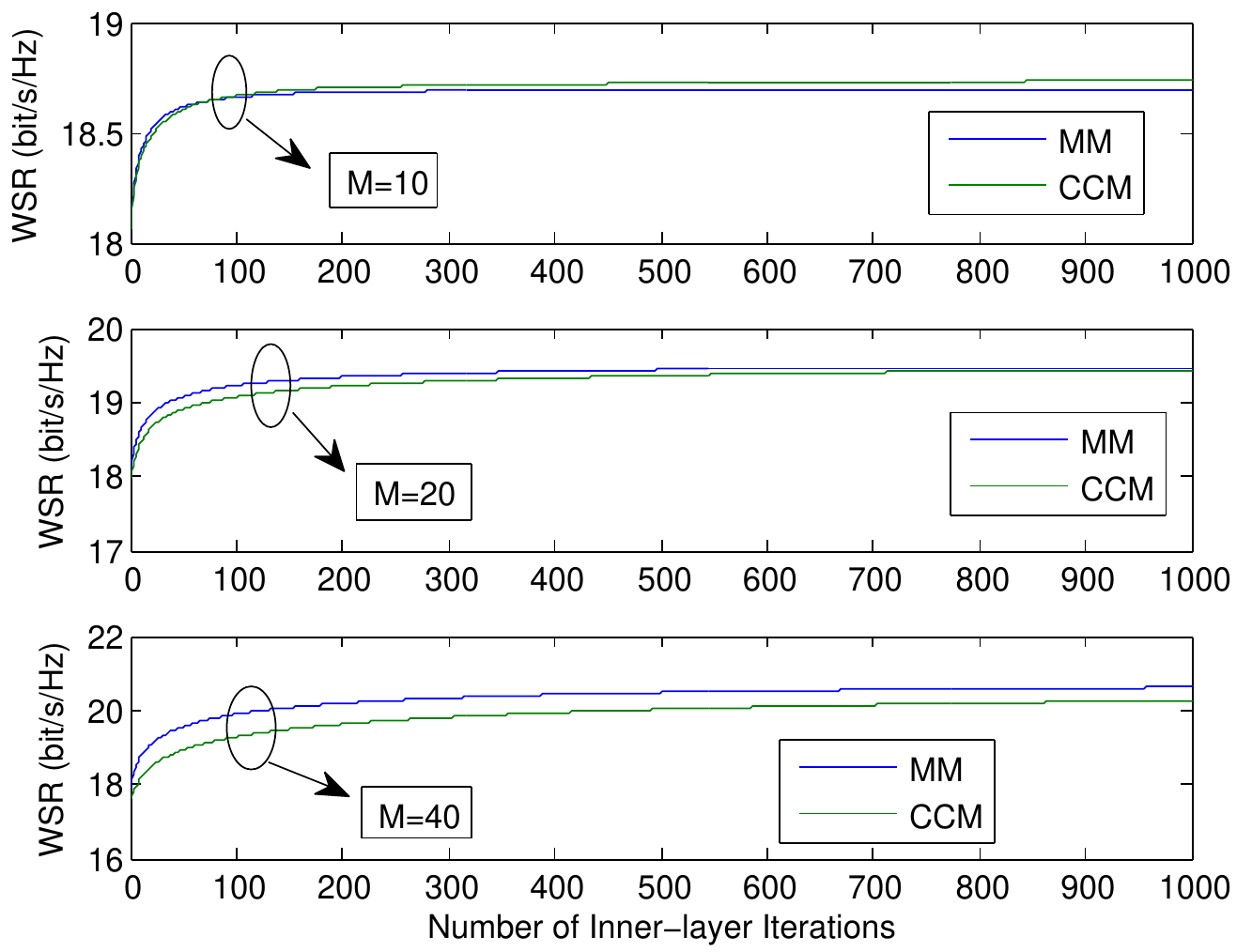}\vspace{-0.6cm}
\caption{Convergence behaviour of the MM and CCM algorithm.}
\label{fig5}
\end{minipage}\vspace{-0.7cm}
\end{figure}

We first study the convergence behaviour of the BCD algorithm in Algorithm \ref{bcd}. Fig.~\ref{fig4} shows the WSR versus the number of iterations for various number of phase shifts, i.e., for $M=10, 20$ and $40$. Both the BCD-MM and BCD-CCM algorithms are tested. It can be observed from this figure that both the BCD-MM and BCD-CCM have a very similar  convergence speed and converged value. Having more phase shifts leads to a slightly slower convergence speed. This is due to the fact that more optimization variables are involved, and more iterations are required for convergence. However, for different values of $M$, the proposed algorithms converge within 200 iterations, which confirm the practical benefits of our algorithms.

\subsubsection{Convergence behaviour of the MM and CCM algorithms}

In each iteration of the BCD algorithm, we have to use the MM or CCM algorithm for finding the phase shifts of the IRS. Fig.~\ref{fig5} shows the convergence performance of the MM and CCM algorithms for the first iteration of the BCD algorithm. It can be seen from Fig.~\ref{fig5} that the MM algorithm converges a little faster than the CCM algorithm, which implies having a lower computational complexity for the MM algorithm based on the complexity analysis of Table \ref{tabzero}. As expected, the number of iterations required for the convergence of the two algorithms increase with the number of phase shifts, since more variables have to be optimized. For different values of $M$, the MM algorithm and CCM algorithm may converge to different values. However, as seen from Fig.~\ref{fig4}, the final WSR value obtained by the BCD algorithm by using different algorithms to update the phase shifts is almost the same.

We then compare our proposed algorithms to the following benchmark schemes:
\begin{enumerate}
  \item  \textbf{RandPhase}: We assume that the phase for each reflection element is uniformly and independently generated from $[0,2\pi]$. We only have to optimize the TPC matrices, which can be obtained by skipping Step 5 of the BCD algorithm.
  \item \textbf{No-IRS}: Set the IRS related channel matrices to zero matrices, i.e., ${\bf{H}}_{r,l,k}={\bf{0}}$,  ${{\bf{G}}_{n,r}}={\bf{0}},\forall n,l,k$. Then, use the BCD algorithm to find the optimal TPC matrices by removing Step 5 for the phase shift update.
\end{enumerate}
%
%

\subsubsection{Impact of the Number of Phase Shifts}

\begin{figure}
\begin{minipage}[t]{0.495\linewidth}
\centering
\includegraphics[width=2.6in]{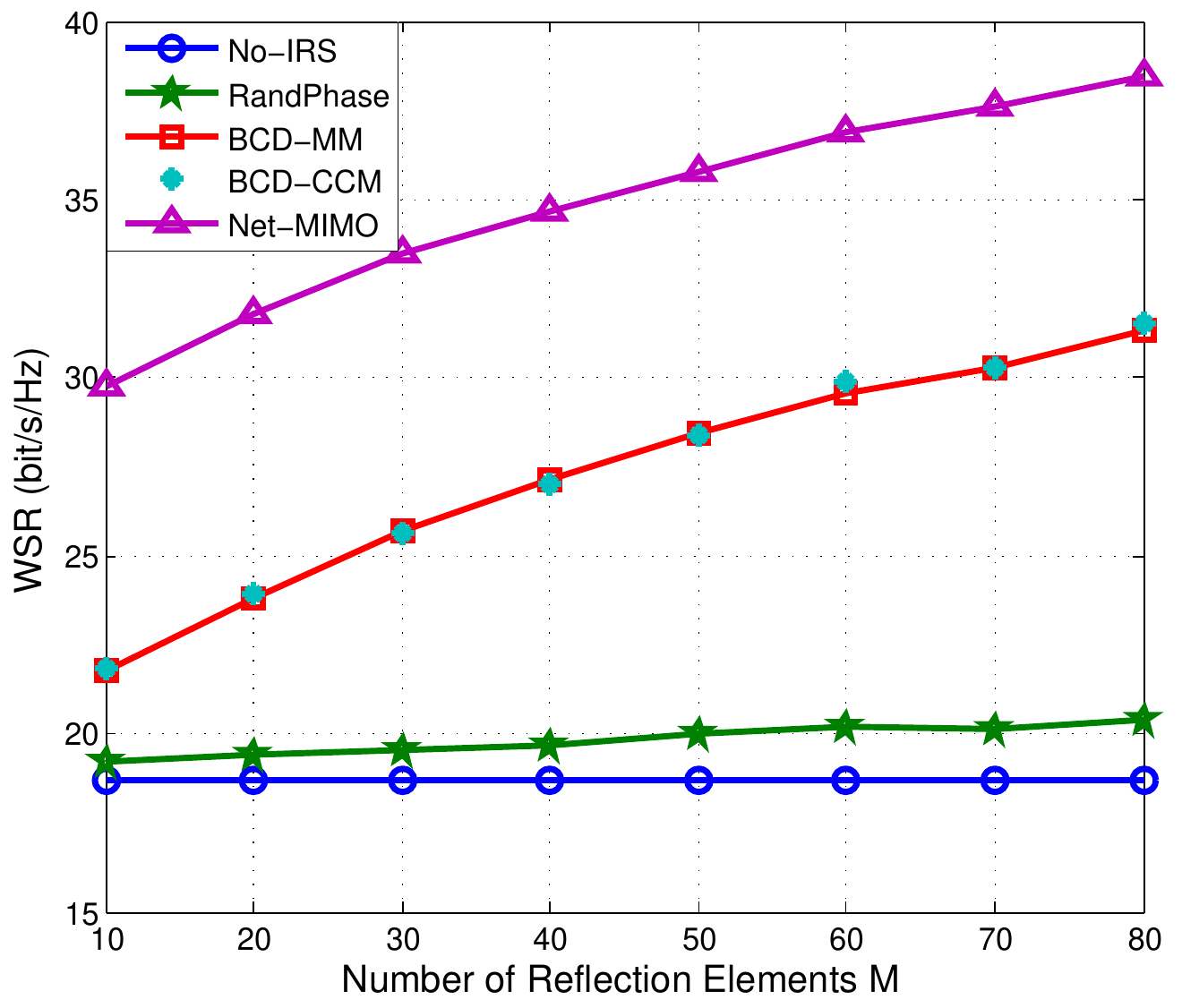}\vspace{-0.6cm}
\caption{Achievable WSR versus the number of phase shifts $M$.}
\label{fig7}
\end{minipage}%
\hfill
\begin{minipage}[t]{0.495\linewidth}
\centering
\includegraphics[width=2.6in]{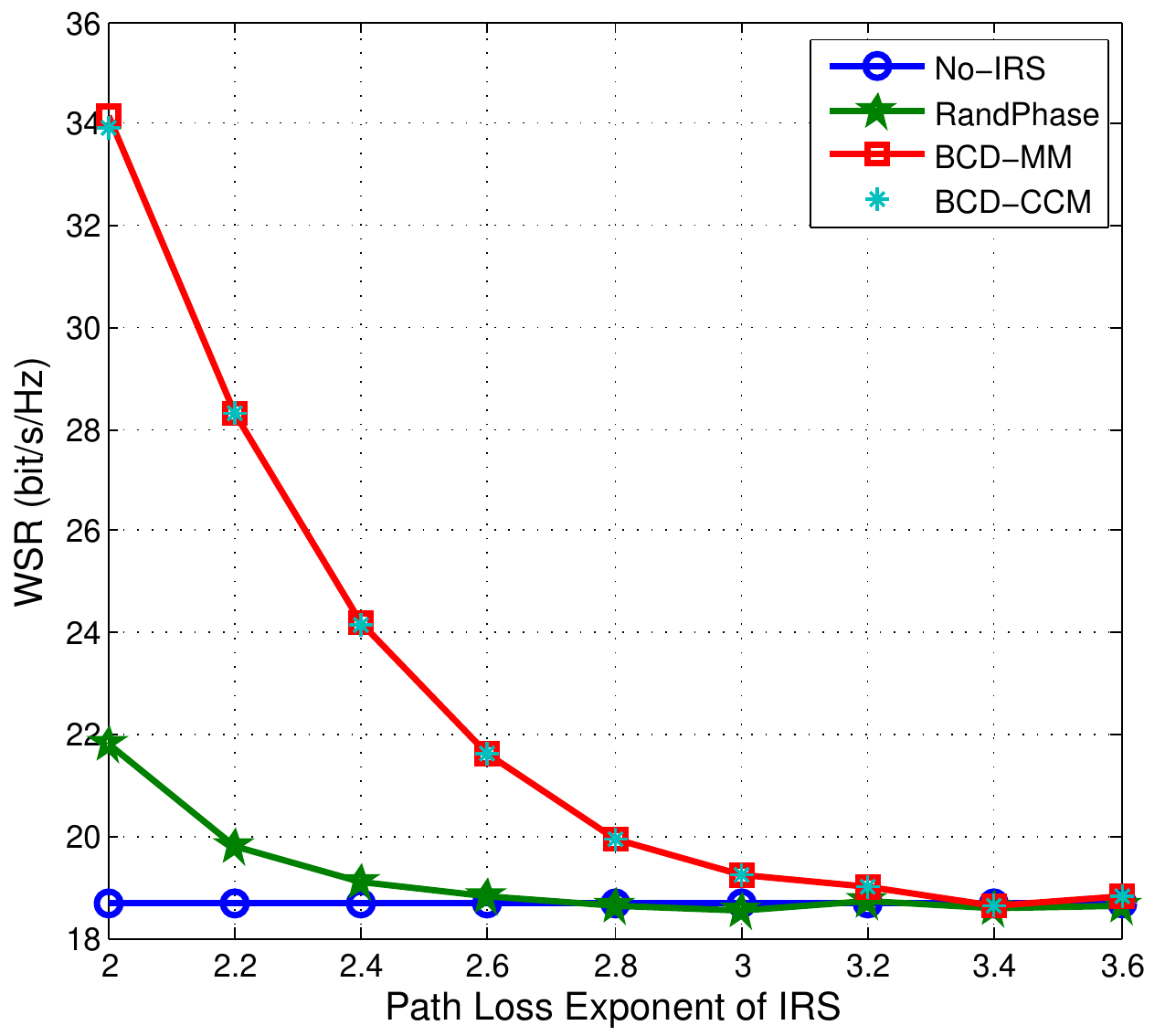}\vspace{-0.6cm}
\caption{Achievable WSR versus   IRS-related path loss exponent.}
\label{fig8}
\end{minipage}\vspace{-0.7cm}
\end{figure}

Fig.~\ref{fig7} compares the WSR performance of various algorithms versus the number of phase shifts $M$. The performance of Network MIMO scheme (with legend `Net-MIMO') proposed in Section \ref{joshgtro} is also compared.  We can  observe that both the BCD-MM algorithm and BCD-CCM algorithm have similar performances over the entire range of $M$, and both of them significantly outperform the other two benchmark schemes.  The performance gain becomes quite pronounced upon increasing $M$. Specifically, when $M=10$, the performance gain over the No-IRS is only 2 bit/s/Hz, while the performance gain increases up to 13 bit/s/Hz when $M=80$. This is mainly attributed to two reasons. Firstly, the signal power received  at the IRS can be enhanced by increasing $M$, leading to a higher array gain. On the other hand, by appropriately designing the phase shifts, the reflected signal power received by the users increases with  $M$. Hence, the proposed IRS-assisted system can exploit not only the array gain, but also the reflecting beamforming gain at the IRS. More importantly, the IRS is a passive reflection device, hence installing more passive reflecting elements is both energy-efficient and economical since the IRS does not require active radio frequency chains and power amplifiers as in conventional transmitters. These results demonstrate that introducing IRSs into wireless communications   enhances the system performance, and it is a promising technique for future networks. It is seen that the performance of the RandPhase algorithm is slightly better than that of the No-IRS scheme. This is because the reflected signals have not been carefully beamed towards the receivers. By contrast, for the proposed algorithms, both the direct signals and reflected signals are superposed more constructively, while the multicell interference signals are added  destructively. As expected, the WSR achieved by the Network MIMO is significantly higher than that of the  system studied in Section \ref{system} (denoted as `Coordinated beamforming'). However, this performance gain is attained at the cost of the heavy information exchange associated with Network MIMO, where the data streams of all the users should be exchanged. By contrast, only the CSI has to be shared among the BSs, the amount of which is much lower than that of the data.

\subsubsection{Impact of the IRS-related Path Loss Exponent}

In the above examples, the path loss exponents of the IRS-related links is set as ${\alpha _{{\rm{IRS}}}}= 2.2$, since we assume that  the location of the IRS can be appropriately chosen for ensuring that a free space  BS-IRS link and IRS-user link can be established. However, in some practical scenarios, it may not be feasible to find such ideal places. Hence, it is intriguing to investigate the performance gain that can be achieved by our proposed algorithms when the IRS-related links experience rich scattering fading with higher value of ${\alpha _{{\rm{IRS}}}}$. To this end, we plot Fig.~\ref{fig8} to show the impact of the IRS-related path-loss exponent. As expected, the WSR achieved by the proposed algorithms decreases upon increasing ${\alpha _{{\rm{IRS}}}}$, and finally converges to the same WSR as achieved by the No-IRS scheme. This is because upon increasing ${\alpha _{{\rm{IRS}}}}$, the signal attenuation associated with the IRS-related links becomes larger, and the signal received from the IRS is weaker, hence more negligible. However, when  ${\alpha _{{\rm{IRS}}}}$ is very small, significant performance gains can be achieved by our proposed algorithms over the No-IRS scheme. For example, for a free-space channel associated with  ${\alpha _{{\rm{IRS}}}}=2$, the performance gain is up to 14.5 bit/s/Hz. Hence, for multicell systems, the performance gain of IRS-assisted systems may be attributed to the favourable channel conditions of the BS-IRS link and IRS-user link. This provides an important engineering design insight, where the IRS should be deployed in an obstacle-free scenario, such as the ceiling for indoor use or advertisement panels for outdoor use. Otherwise, the performance gain brought about by the IRS is marginal.  Fig.~\ref{fig8} also shows that if the phase shifts are not optimized, the performance  of an IRS-aided system may  even be worse than that operating without the IRS, i.e. the WSR achieved by the RandPhase algorithm is equal to or lower than that of the No-IRS scheme. This emphasizes  the importance of jointly optimizing  the TPC matrices and the phase shifts at the IRS.

\subsubsection{Impact of the IRS Location}

\begin{figure}
\begin{minipage}[t]{0.495\linewidth}
\centering
\includegraphics[width=2.6in]{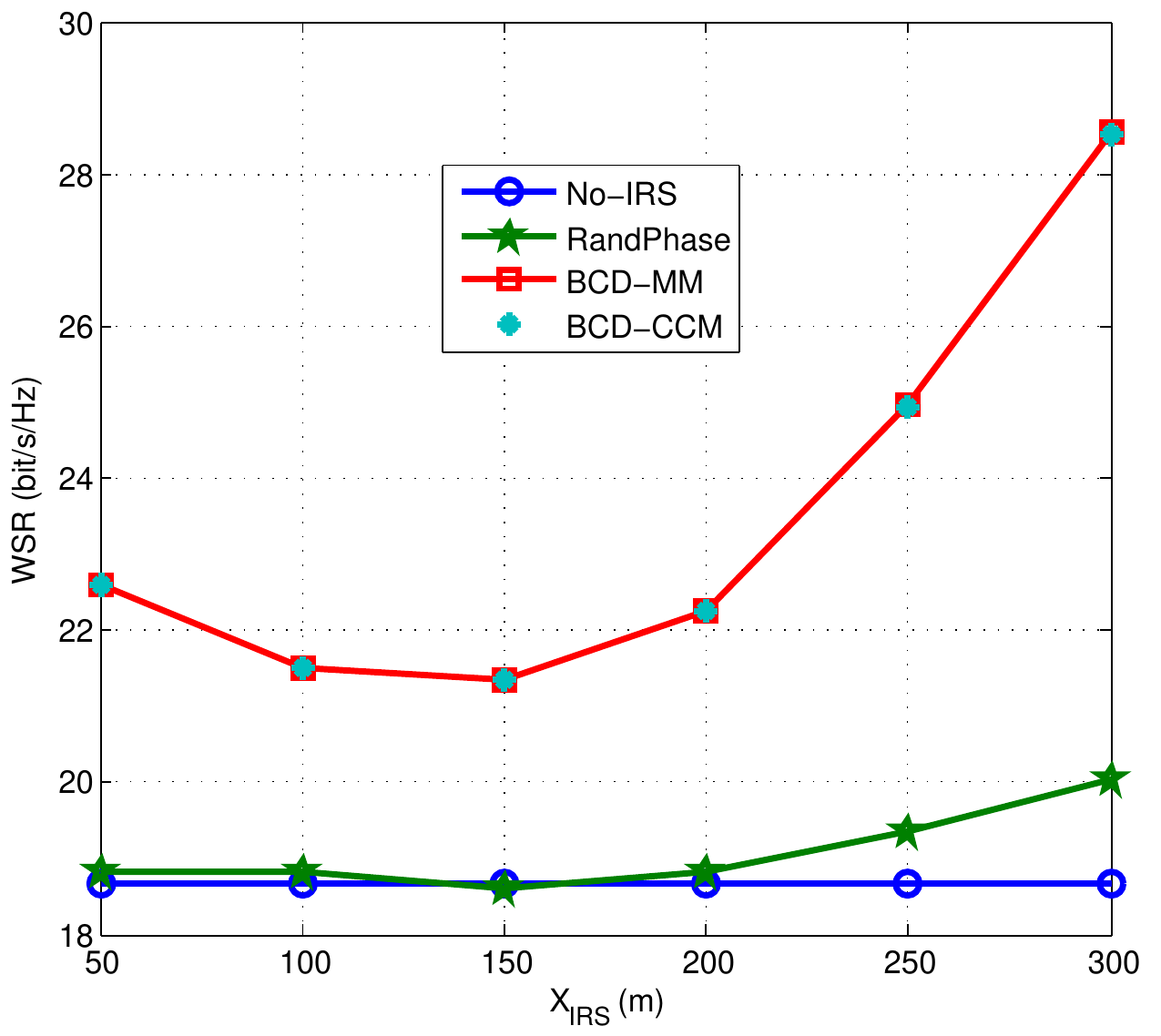}\vspace{-0.6cm}
\caption{Achievable WSR versus the location of the IRS $x_{\rm{IRS}}$.}
\label{fig9}
\end{minipage}%
\hfill
\begin{minipage}[t]{0.495\linewidth}
\centering
\includegraphics[width=2.6in]{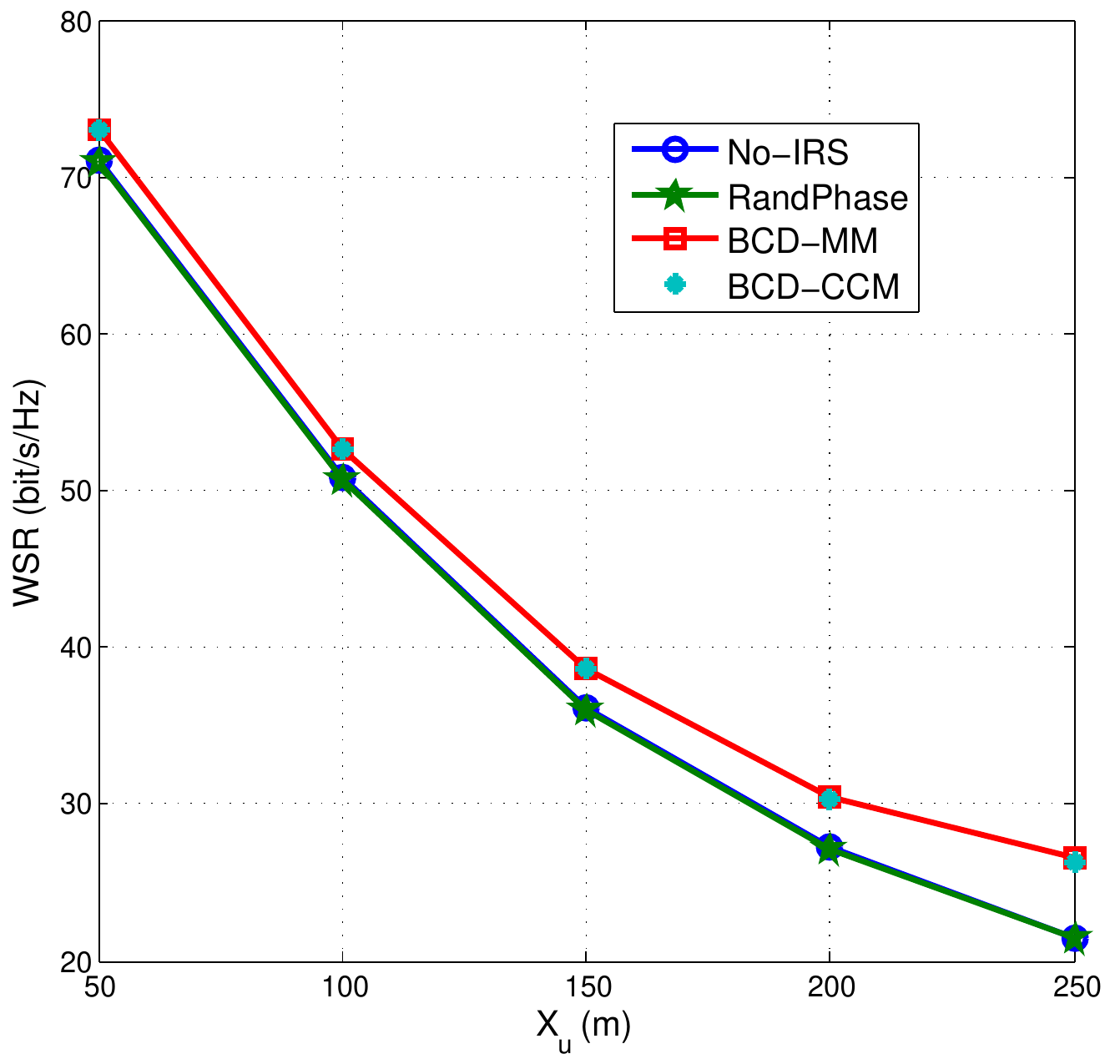}\vspace{-0.6cm}
\caption{Achievable WSR versus the location of the UE $x_u$.}
\label{fig10}
\end{minipage}\vspace{-0.7cm}
\end{figure}

Denote the coordinate of the IRS as $(x_{\rm{IRS}},0)$. In Fig.~\ref{fig9}, we study the impact of the IRS location by moving the IRS from $x_{\rm{IRS}}=50$ m (cell center of the first cell) to $x_{\rm{IRS}}=300$ m (cell boundary). It may be observed again that both the proposed algorithms achieve the similar performance, and drastically improves the WSR performance over the other benchmark schemes. It is interesting to observe that the WSR achieved by the proposed algorithms first decreases with $x_{\rm{IRS}}$ ($50$ m$ <x_{\rm{IRS}}<150$ m), and then increases for $x_{\rm{IRS}}>150$ m. This becomes plausible upon considering a special case, where the IRS lies on the line between the BS and the user central point. Let us denote  the distance between the BS and the IRS by $d$, and that  between the BS and the user central point by $D$.  By ignoring the small-scale fading, the large-scale channel gain of the combined channel from the IRS may be approximated by
\begin{equation}\label{xadefwre}
 {\rm{PL}}_{\rm{IRS}}=2{\rm{P}}{{\rm{L}}_0} - 10\alpha_{\rm{IRS}} {{\log }_{10}}\left( d \right) -10\alpha_{\rm{IRS}} {{\log }_{10}}\left( D-d \right),
\end{equation}
which achieves its minimum value at $d^\star=D/2$. Hence, the combined channel gain  achieves its minimum value when the IRS is located at the middle point, which is consistent with the simulation results of Fig.~\ref{fig9}. Due to the strong BS-IRS link, the WSR performance gain achieved by our proposed algorithms over the No-IRS is 4 bit/s/Hz at $x_{\rm{IRS}}=50$ m. However, this performance gain doubles when the IRS moves to the boundary of these two cells. This performance is partly due to the favourable IRS-user channel link. The other important reason is that we can optimize the phase shifts of the IRS to make the equivalent channel spanning from the inter-cell BS to  the users approach zero matrices. Specifically, we can optimize ${\bm{\Phi}}$ to let  ${\bf{\bar H}}_{n,l,k}, n\ne l$ approach zero matrices. This  alleviates the severe inter-cell interference for the cell-edge users, which significantly enhances the system performance. Additionally, deploying the IRS at the cell center for Cell 1 is only beneficial for the users in Cell 1, while  all the users will benefit from the IRS, when positioning it at the cell boundary. This means that for multicell communication systems, significant performance gains can be obtained when the IRS is employed at the cell boundary, which mitigates the inter-cell interference. Furthermore, the phase shifts should be carefully designed. Otherwise, the performance may in fact become inferior to that without IRS, e.g., $x_{\rm{IRS}}=150$ m.

\subsubsection{Impact of the User Location}

In Fig.~\ref{fig10}, we compare the WSR achieved by all schemes versus the horizontal distance between BS 1 and the first circle central point, i.e., $x_u$. Since the users are randomly positioned in this circle, this is equivalent to varying the locations of the users. It is again observed that the proposed  algorithms achieve almost the same performance and achieve superior performance over the other two benchmark schemes. Additionally, the performance gap increases with  $x_u$, because the users   receive strong reflected signals from the IRS, when the users  approach the cell edge. This means that the IRS  mitigates the inter-cell interference.

\begin{figure}
\begin{minipage}[t]{0.495\linewidth}
\centering
\includegraphics[width=2.6in]{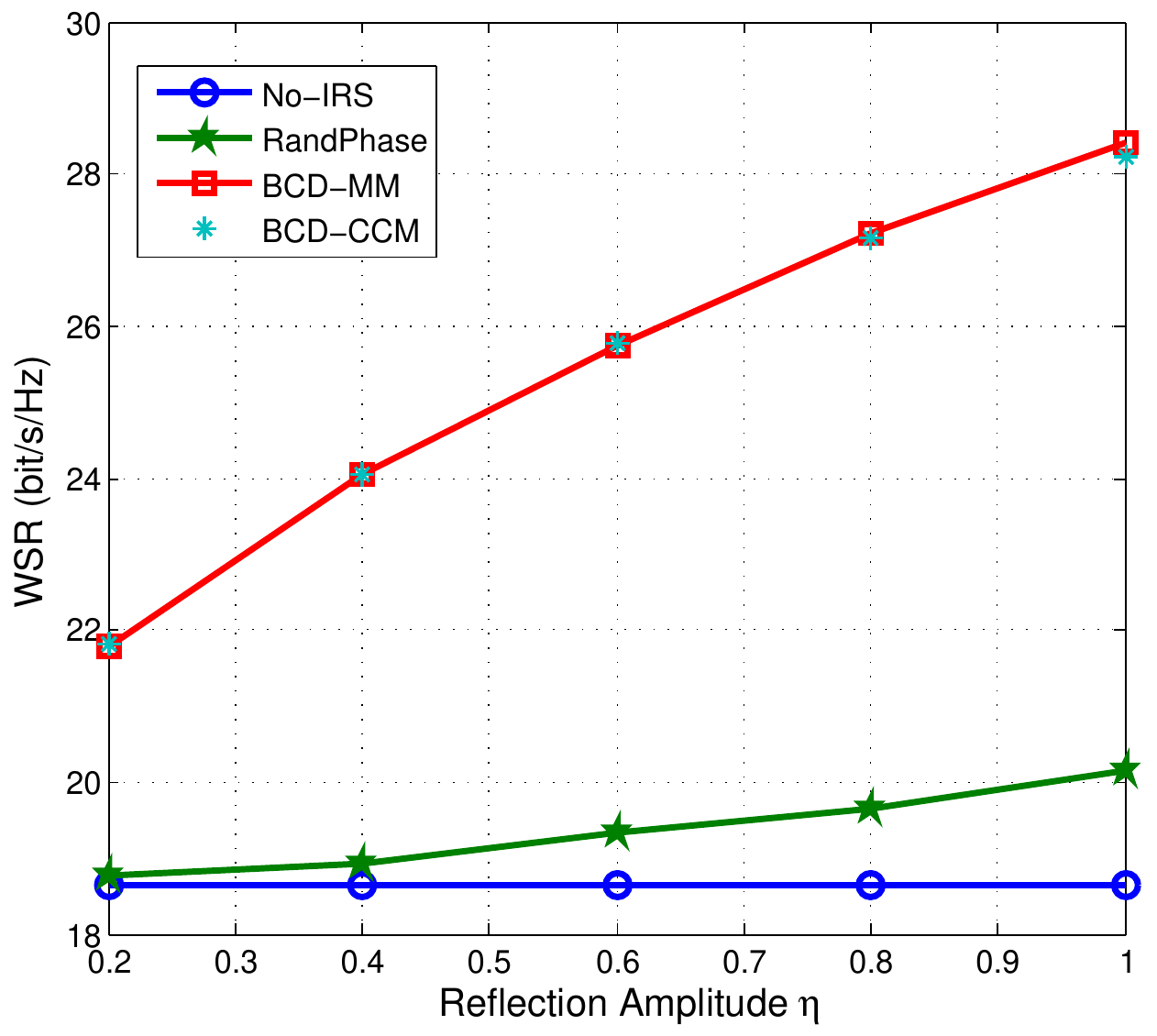}\vspace{-0.6cm}
\caption{ Achievable WSR versus the reflection amplitude $\eta$. }
\label{fig11}
\end{minipage}%
\hfill
\begin{minipage}[t]{0.495\linewidth}
\centering
\includegraphics[width=2.6in]{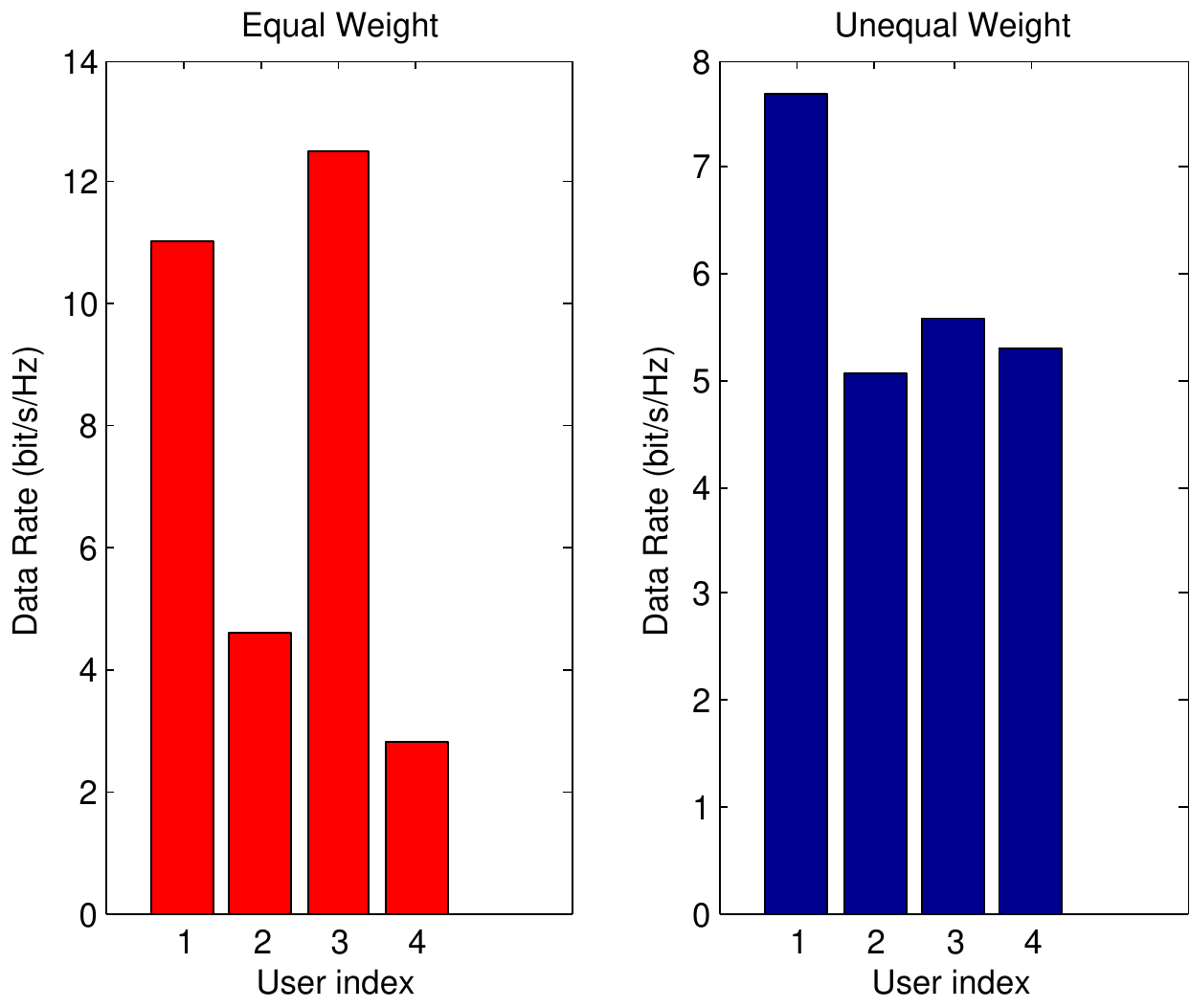}\vspace{-0.6cm}
\caption{ Individual  data rate under two sets of weights. }
\label{fig12}
\end{minipage}\vspace{-0.7cm}
\end{figure}

\subsubsection{ Impact of the Reflection Amplitude }

 Due to the absorption and parasitic reflection of the phase shifters, there may be a signal power loss at the IRS. Then, in Fig.~\ref{fig11}, we study the impact of the reflection amplitude on the system performance. Specifically, the  phase-shift matrix of the IRS is rewritten as ${\bm{\Phi}}  = \eta{\rm{diag}}\left\{ {{e^{j{\theta _1}}}, \cdots ,{e^{j{\theta _m}}}, \cdots ,{e^{j{\theta _M}}}} \right\}$, where the reflection amplitudes of all the elements are the same as $\eta$. As expected, the WSR achieved by the IRS-aided scheme increases with $\eta$ due to the reduced power loss. The reflection amplitude has a substantial impact on the system performance. Specifically, when $\eta$ increases from 0.2 to 1, the WSR increases by about 6 bit/s/Hz.

\subsubsection{ Impact of the Weights }

 As mentioned in our problem formulation, the weights can be used for controlling the fairness among the users. To be more explicit, we provide an example for illustrating this point. For clarity, the index of the $j$th user in the $i$th cell
is denoted as $2(i-1)+j$. For example, the index  of the second user in the second cell is 4.
The coordinates for the four users (two in each cell) are respectively given by $(100,0),(250,0),(350,0)$ and $(500,0)$, which indicate that the first user is closer to BS 1 than the second user, and the third user is closer to BS 2 than the fourth user.
Two sets of weights are tested: 1) $\omega_{k}=0.5, \forall k$; 2) $\omega_{1}=0.15, \omega_{2}=0.85, \omega_{3}=0.3, \omega_{4}=0.7$. In Fig.~\ref{fig12}, the individual data rates achieved under two sets of weights are illustrated. For the case of the equal weights, the first user and the third user have higher data rate than the other two users, since they are closer to the BSs. To guarantee rate-fairness amongst the users, for the case of unequal weights, a more balanced data rate distribution can be achieved by assigning higher weights to the users having low channel gains.
\begin{figure}
\centering
\includegraphics[width=2.8in]{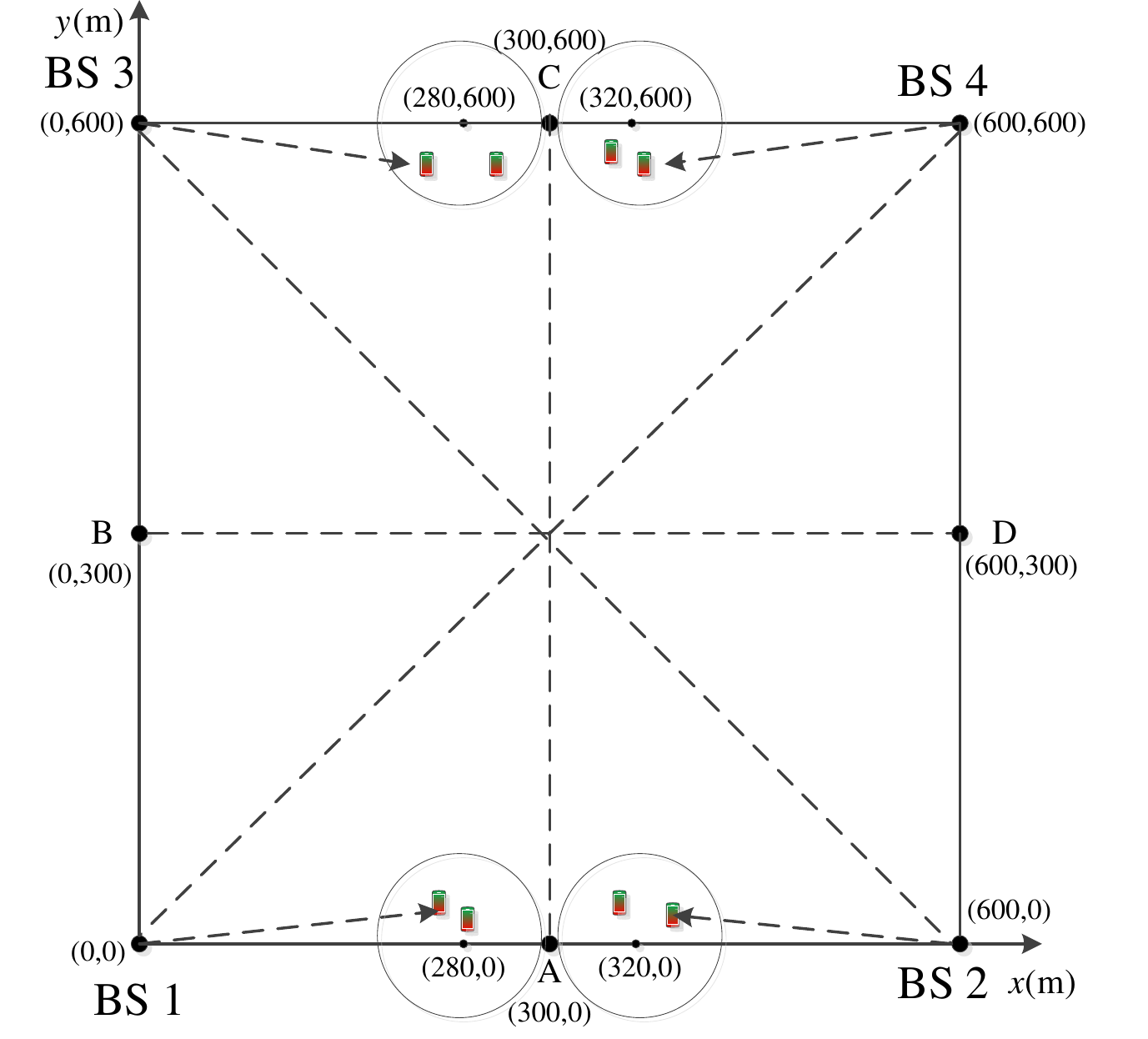}\vspace{-0.2cm}
\caption{ The simulated four-cell IRS-aided MIMO communication scenario. }\vspace{-0.5cm}
\label{fig13}
\end{figure}
\subsection{ Four-cell Scenario }
 Finally, in order to study the beneficial impact of IRS deployment on the system's performance, we consider the four-cell scenario of Fig.~\ref{fig13}, where the coordinates of the four BSs are given by $(0,0),(600,0),(0,600)$ and $(600,600)$, respectively. Additionally, the coordinates of the user distribution center in the four cells are $(280,0),(320,0),(280,600)$ and $(320,600)$, respectively. The circle radius is also 20 m. Four points (i.e., A,B,C,D) are located at the middle of the corresponding two BSs. The number of antennas at each BS is set to 2, and each cell has three users.

\begin{figure}
\begin{minipage}[t]{0.495\linewidth}
\centering
\includegraphics[width=2.6in]{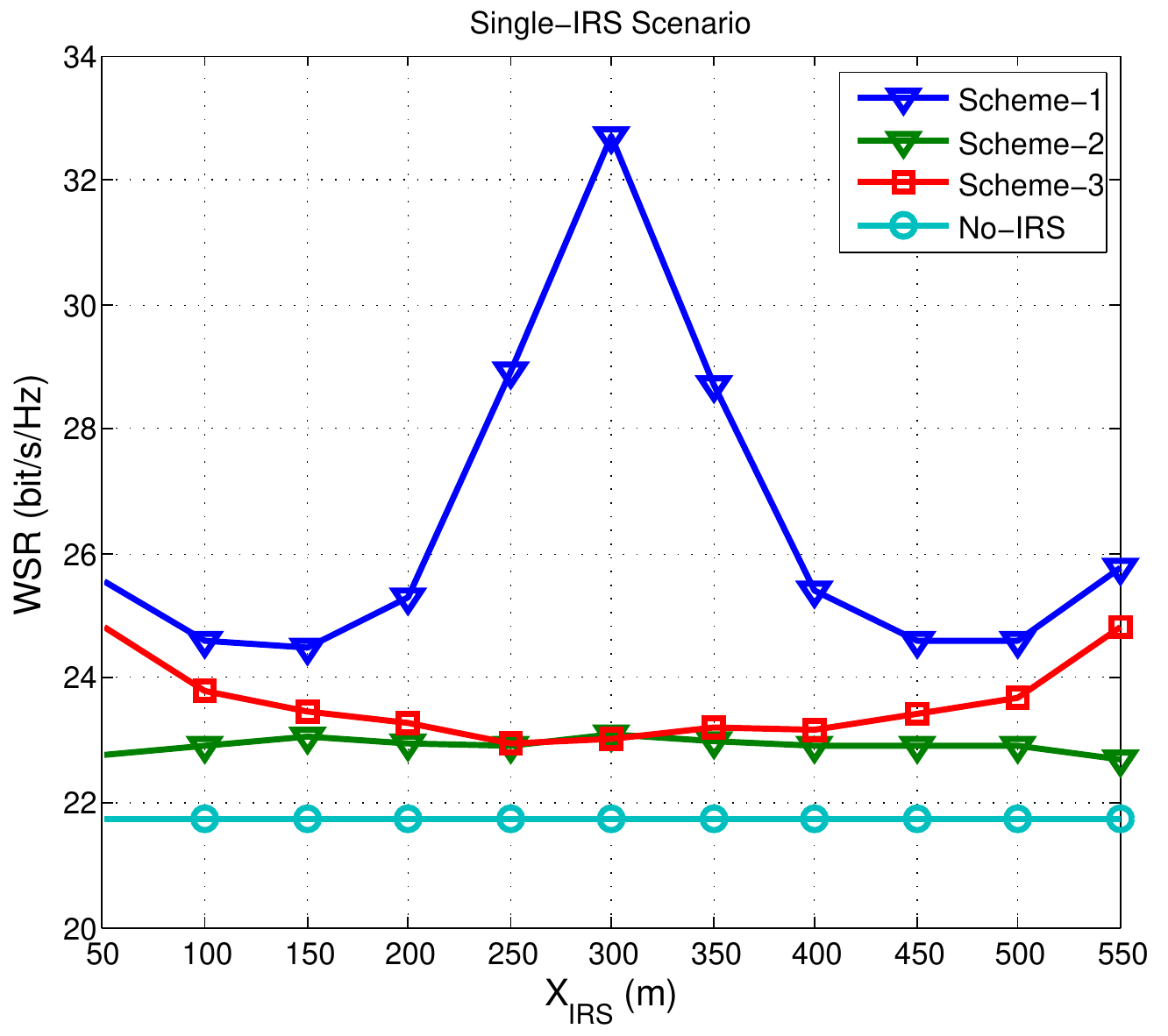}\vspace{-0.6cm}
\caption{ Achievable WSR versus various IRS deployment schemes for single-IRS case. }
\label{fig14}
\end{minipage}%
\hfill
\begin{minipage}[t]{0.495\linewidth}
\centering
\includegraphics[width=2.6in]{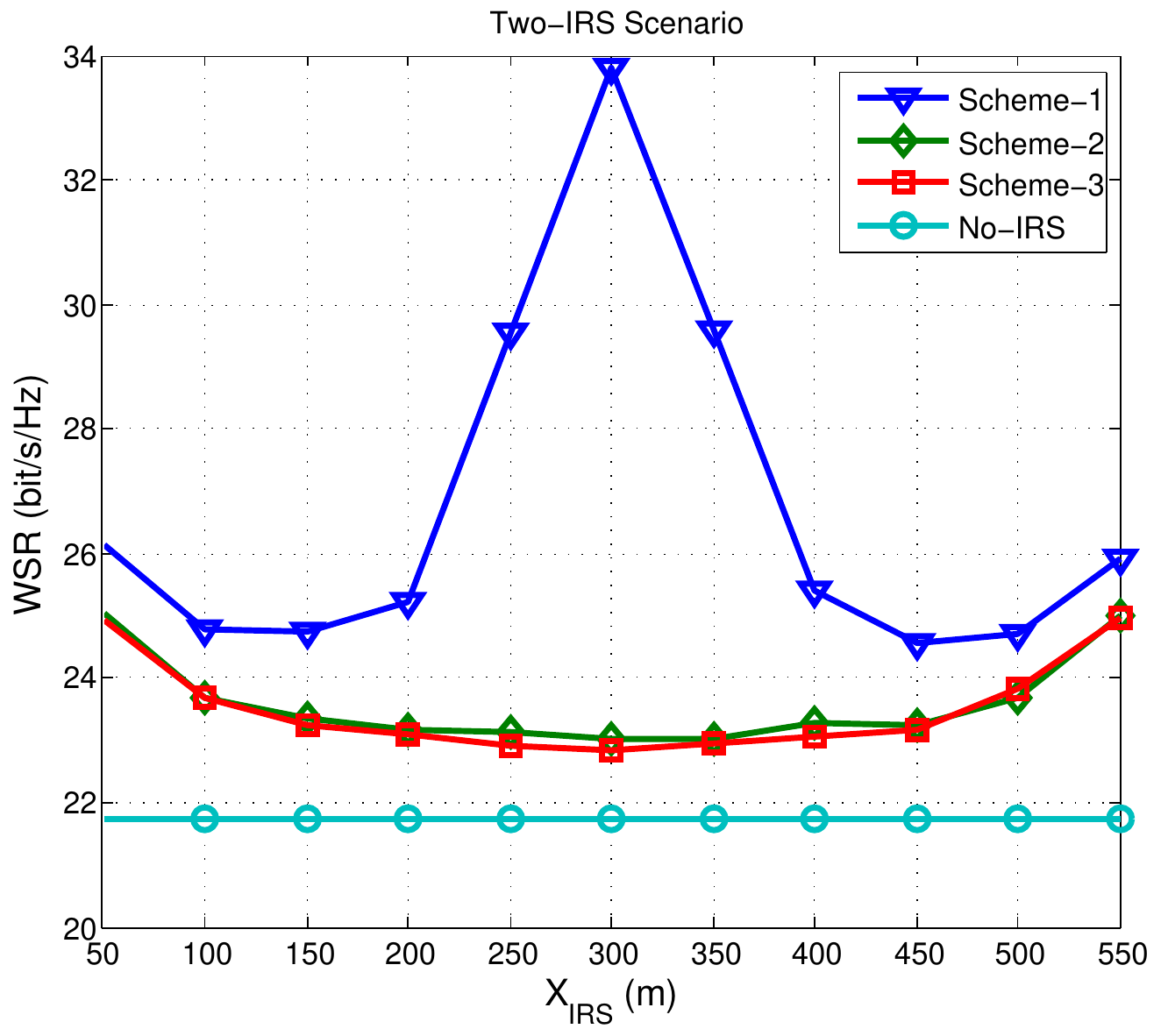}\vspace{-0.6cm}
\caption{ Achievable WSR versus various IRS deployment schemes for two-IRS case. }
\label{fig15}
\end{minipage}\vspace{-0.7cm}
\end{figure}
\subsubsection{ Single-IRS Case}
 We first study the single-IRS scenario, when the number of phase shifts at the IRS is 50.  Three IRS schemes are considered: 1) \textbf{Scheme-1}: As in the case of two-cell scenario, the IRS moves from BS 1 to BS 2; 2) \textbf{Scheme-2}: The IRS moves from point B to point D; 3) \textbf{Scheme-3}: The IRS moves from BS 3 to BS 2.
The WSR achieved by various schemes is shown in Fig.~\ref{fig14}. It is seen from this figure that Scheme 1 achieves its maximum WSR at the cell boundary point with $x_{\rm{IRS}}=300 m$, which implies that the IRS should be deployed at the cell edge to benefit the users in the first and second cells. This conclusion is consistent with that of the two-cell scenario shown in Fig.~\ref{fig9}.
It is also shown that Scheme-1 has the best performance for any locations of the IRS.
The reason may be that the IRS in Scheme-1 is more close to  the first and second users. However, the users in the third and fourth cells are far away from the IRS in Scheme-1,   thus the benefits of the IRS for these users are marginal. This motivates the deployment of more IRSs in the system. Again, the WSR achieved by the various schemes is higher than that without IRS, which demonstrates the benefits of installing IRSs in multicell networks.
\subsubsection{ Two-IRS Case }
 In this case, two IRSs are deployed, each of which has 25 phase shifts. Hence, the total number of phase shifters is equal to that of the single-IRS case. Three schemes are considered: 1) \textbf{Scheme-1}: IRS 1 moves from BS 1 to BS 2, and IRS moves from BS 3 to BS 4; 2) \textbf{Scheme-2}: IRS 1 moves from BS 1 to BS 4, and IRS 2 moves from BS 3 to BS 2; 3) \textbf{Scheme-3}: IRS 1 moves from point B to point D, and IRS 2 moves from point C to point A. The WSR achieved by various schemes is shown in Fig.~\ref{fig15}. Similar trends have been observed to the single-IRS case. For example, Scheme-1 performs the best, and achieves its highest WSR when the IRS is located at points A and C, respectively. The reason is that the IRSs are closer to the users in these two points. By comparing Fig.~\ref{fig14} and Fig.~\ref{fig15}, when $X_{\rm{IRS}}=300 m$, the WSR of the Scheme-1 in the two-IRS scenario is higher than that in the single-IRS scenario, which means that the distributed IRS deployment is more beneficial than centralized deployment. In general, the number of IRSs depends on the number of user clusters. It is expected that in the vicinity of each user cluster, there is at least one IRS.

\section{Conclusions}\label{conclu}

In this paper, we have enhanced the cell-edge user performance of multicell communication systems by employing an IRS at the cell boundary. Specifically, by carefully tuning the phase shifts, the inter-cell interference reflected by IRS can be added destructively to that directly  received from  the adjacent BSs, which  alleviates the inter-cell interference received by the cell-edge users. We studied the WSR maximization problem by jointly optimizing the active TPC matrices at the BSs and passive shifts at the IRSs, while guaranteeing each BS's power constraint and unit-modulus constraint at the IRS. To tackle this non-convex problem, the BCD algorithm was used for optimizing them in an alternating manner. The optimal TPC matrices were obtained in closed form, and a pair of   efficient algorithms were provided for solving the challenging phase shift optimization problem. Our simulation results verified that the proposed algorithms  achieve significant performance gains over their conventional counterpart operating without incorporating an IRS. Furthermore, the location of IRS should be carefully chosen to guarantee a favourable  BS-IRS link and IRS-user link.  When the IRSs are deployed in the vicinity of user clusters,  distributed IRS deployment is shown to be advantageous over the centralized deployment.

%
%


\
\





\vspace{-0.5cm}
\bibliographystyle{IEEEtran}
\bibliography{myre}


\end{document}